\documentclass[aps, prl, amsmath, floats,floatfix, twocolumn, superscriptaddress]{revtex4-1}

\usepackage[normalem]{ulem}
\usepackage{amssymb}
\usepackage{amsmath}
\usepackage{verbatim}
\usepackage{mathrsfs}
\usepackage{amsfonts}
\usepackage{latexsym}
\usepackage{epsfig}
\usepackage{color}
\usepackage{graphicx,subfigure}
\usepackage{units}
\usepackage{slashbox}
\usepackage{envmath}
\usepackage{natbib}
\usepackage{hyperref}
\usepackage[utf8]{inputenc}
\usepackage{bm}

\begin{document}


\definecolor{orange}{rgb}{0.9,0.45,0}

\newcommand{\re}{\mbox{Re}}
\newcommand{\im}{\mbox{Im}}

\newcommand{\tf}[1]{\textcolor{red}{#1}}
\newcommand{\nsg}[1]{\textcolor{blue}{#1}}
\newcommand{\ch}[1]{\textcolor{red}{CH: #1}}
\newcommand{\fdg}[1]{\textcolor{orange}{FDG: #1}}
\newcommand{\pcd}[1]{\textcolor{magenta}{#1}}
\newcommand{\mz}[1]{\textcolor{cyan}{[\bf MZ: #1]}}

\def\CovDev{D}
\def\Res{{\mathcal R}}
\def\Gammaflat{\hat \Gamma}
\def\metricflat{\hat \gamma}
\def\Dflat{\hat {\mathcal D}}
\def\part_n{\partial_\perp}

\def\Lie{\mathcal{L}}
\def\A{\mathcal{X}}
\def\Aphi{\A_{\phi}}
\def\hAphi{\hat{\A}_{\phi}}
\def\E{\mathcal{E}}
\def\Ham{\mathcal{H}}
\def\M{\mathcal{M}}
\def\R{\mathcal{R}}
\def\p{\partial}

\def\hg{\hat{\gamma}}
\def\hA{\hat{A}}
\def\hD{\hat{D}}
\def\hE{\hat{E}}
\def\hR{\hat{R}}
\def\hcA{\hat{\mathcal{A}}}
\def\hDelt{\hat{\triangle}}

\def\be{\begin{equation}}
\def\ee{\end{equation}}

\renewcommand{\t}{\times}

\long\def\symbolfootnote[#1]#2{\begingroup%
\def\thefootnote{\fnsymbol{footnote}}\footnote[#1]{#2}\endgroup}


\title{Exotic Compact Objects and the Fate of the Light-Ring Instability}

    \author{Pedro V. P. Cunha}
\affiliation{Departamento  de  Matem\'{a}tica  da  Universidade  de  Aveiro  and  Centre  for  Research  and  Development in  Mathematics  and  Applications  (CIDMA),  Campus  de  Santiago,  3810-183  Aveiro,  Portugal}

    \author{Carlos Herdeiro}
\affiliation{Departamento  de  Matem\'{a}tica  da  Universidade  de  Aveiro  and  Centre  for  Research  and  Development in  Mathematics  and  Applications  (CIDMA),  Campus  de  Santiago,  3810-183  Aveiro,  Portugal}

  \author{Eugen Radu}
\affiliation{Departamento  de  Matem\'{a}tica  da  Universidade  de  Aveiro  and  Centre  for  Research  and  Development in  Mathematics  and  Applications  (CIDMA),  Campus  de  Santiago,  3810-183  Aveiro,  Portugal}

\author{Nicolas Sanchis-Gual}
\affiliation{Departamento de Astronom\'{i}a y Astrof\'{i}sica, Universitat de Val\`{e}ncia,
Dr. Moliner 50, 46100, Burjassot (Val\`{e}ncia), Spain}
	\affiliation{Departamento  de  Matem\'{a}tica  da  Universidade  de  Aveiro  and  Centre  for  Research  and  Development in  Mathematics  and  Applications  (CIDMA),  Campus  de  Santiago,  3810-183  Aveiro,  Portugal}


\date{July 2022}


\begin{abstract} 
Ultracompact objects with light-rings (LRs) but without an event horizon could mimic black holes (BHs) in their strong gravity phenomenology. But are such objects dynamically viable? Stationary and axisymmetric ultracompact objects that can form from smooth, quasi-Minkowski initial data must have at least one \textit{stable} LR, which has been argued to trigger a spacetime \textit{instability}; but its development and fate  have been unknown. Using fully non-linear numerical evolutions of ultracompact bosonic stars free of any other known instabilities and introducing a novel adiabatic effective potential technique, we confirm the LRs triggered instability, identifying two possible fates: migration to non-ultracompact configurations or collapse to BHs. In concrete examples we show that typical migration/collapse time scales are not larger than $\sim 10^3$ light-crossing times, unless the stable LR potential well is very shallow. Our results show that the LR instability is effective in destroying horizonless ultracompact objects that could be plausible BH imitators.

\end{abstract}


\maketitle

\vspace{0.8cm}

\noindent{\bf {\em Introduction.}} 
A few years after the first gravitational wave detection from a collision of two black holes (BHs)~\cite{Abbott2016} and the first image of a BH resolving its horizon scale structure~\cite{Akiyama:2019cqa}, there is a scientific consensus about the physical reality of BHs. Yet, both the inability to observationally proof the \textit{``BH hypothesis"}~\cite{Berti:2015itd,Barack:2018yly,LISA:2022kgy,Herdeiro:2022yle} and its challenging and far-reaching theoretical consequences~\cite{Penrose:1964wq,Hawking:1976ra,Almheiri:2012rt}, demand a thorough scrutiny of its alternatives. 

In this spirit, a variety of horizonless exotic compact objects (ECOs) have been proposed~\cite{Cardoso:2019rvt}: the \textit{``ECO hypothesis"}. Any putative ECO model must overcome theoretical and observational tests to become a contender. Of special interest are \textit{ultracompact} ECOs (UCOs, for short), $i.e.$ possessing light rings (LRs): planar bound photon orbits that asymptotically flat BHs must possess~\cite{Cunha:2020azh}. UCOs can imitate the (initial) ringdown~\cite{Cardoso:2016rao,Cardoso:2017cqb} and (to some extent) the shadow~\cite{Cunha:2018acu} of BHs making them plausible BH foils \textit{if} they are dynamically viable. 

Conditions for dynamical viability include $(i)$ a plausible formation mechanism, $(ii)$ sufficient stability against the ubiquitous astrophysical perturbations and $(iii)$ embedding in a physically sound effective field theory. It was shown in~\cite{Cunha:2017qtt} that, under generic assumptions, an equilibrium UCO that forms from smooth, quasi-Minkowski initial data, must have at least a pair of LRs, one of which is stable. It has been argued that the existence of stable LRs can trigger a spacetime instability, by trapping massless perturbations that eventually pile up and backreact on the spacetime~\cite{Keir:2014oka,Cardoso:2014sna,Benomio:2018ivy}. The development and fate of this hypothetical generic obstruction to UCOs has, however, been so far unknown.

In this letter we present the development and fate of the LR instability in concrete models, providing evidence that UCOs with a plausible formation mechanism are destroyed in astrophysical timescales,  therefore questioning their viability as BH alternatives.

\noindent{\bf {\em UCOs and stable LRs.}} 
Consider equilibrium, finite ADM mass $M$, asymptotically flat, UCOs described by a stationary, axially-symmetric, circular metric, $g_{\mu\nu}$. In $(t,r,\theta,\varphi)$ coordinates, such that $\partial_t$ and $\partial_\varphi$ are the commuting Killing vectors adapted to stationarity and axisymmetry, respectively, (see~\cite{Cunha:2017qtt,Cunha:2020azh} for details), the effective \textit{dimensionless} potentials 
\begin{equation}
V_\pm = \frac{g_{t\varphi}\mp\sqrt{g_{t\varphi}^2-g_{tt}g_{\varphi\varphi}}}{g_{\varphi\varphi}}\, M \ ,
\label{ep}
\end{equation}
determine LRs (if they exist) as critical points: $\nabla V_\pm=0$. The $\pm$ sign is connected to the LRs rotation sense.
 
One can associate a topological charge, $\mathcal{Q}_i$ to individual LRs~\cite{Cunha:2017qtt,Cunha:2020azh}. The total spacetime topological charge $\mathcal{Q}=\sum_{LRs}\mathcal{Q}_i$ is then invariant under smooth spacetime deformations preserving the asymptotic structure and internal regularity. This means that any horizonless, asymptotically flat, everywhere regular UCO has the same topological LR charge as Minkowski spacetime, $\mathcal{Q}=0$. Since unstable (stable) LRs have $\mathcal{Q}_i=-1$ ($\mathcal{Q}_i=+1$), this implies that in the dynamical formation of UCOs LRs emerge as a stable-unstable pair. This conclusion does not depend on the specific (metric) theory of gravity or on the details of the (incomplete) gravitational collapse~\footnote{In particular, the theorem in~\cite{Cunha:2017qtt} does not require stationarity and axi-symmetry \textit{during} this incomplete collapse.}. It implies that for UCOs with a plausible formation mechanism, the existence of a (Schwarzschild-like) unstable LR - as to imitate BH phenomenology - is accompanied by a potentially dangerous stable LR.

\noindent{\bf {\em Adiabatic effective potential.}} 
We wish to follow the dynamics of UCOs. It has been argued that the instability triggered by stable LRs is non-linear~\cite{Keir:2014oka}. Unveiling its final state, moreover, requires a non-linear analysis. Thus, we shall resort to numerical evolutions of appropriate UCO models, using a 3+1 spacetime split:
\begin{align}
ds^2 &= -N^2\,dt^2 + \gamma_{ij}\left(dx^i + \beta^i\,dt\right)\left(dx^j + \beta^j\,dt\right) \ ,
\label{em}
\end{align}
where $N$ is the lapse function, $\bm \beta$ is the shift, and $\bm\gamma$ is the projected 3-metric on the $t=$constant slice $\Sigma_t$, which uses Cartesian-like coordinates $\{x,y,z\}$. 

To monitor the evolution of LRs we introduce, from the numerical evolution 3+1 data in~\eqref{em}, an \textit{adiabatic effective potential} (AEP) analog to~\eqref{ep} at \textit{each} time step, taking $M$ in~\eqref{em} to be $M_t$, the UCO mass at each time slice $t$, to allow comparisons at different $t$. The AEP carries physical information if the evolution is sufficiently slow and departures for axial-symmetry are mild. The lack of stationarity and axi-symmetry can, moreover, be washed away by appropriate averaging procedures. 

A detailed discussion of the assumptions in constructing the AEP is given in Appendix A; here we state the key ones; 
$(i)$
At each  $t$, approximate Killing vectors $\partial_t$ and  $\partial_\varphi$ exist for the evolving, asymptotically flat and approximately circular UCO. This implies that at each point a $\lambda\in\mathbb{R}$ exists such that $\partial_\varphi = \lambda {\bm\beta}$. $(ii)$  $\partial_\varphi$ is assumed to be tangent to surfaces
with $x^2+y^2\equiv r^2=$ constant (but $r$ needs not be a geometric distance). At each $t$ the coordinate center $\mathcal{O}$: $(x,y)=(0,0)$ is linearly shifted to the UCO center, to account for a possible (slow) drift of the star, which indeed occurs in the examples below. $(iii)$ The evolving UCO is assumed to possess a $\mathbb{Z}_2$ reflection symmetry around the equatorial plane surface $z=0$.

Under these assumptions,  the $(g_{tt},g_{t\varphi},g_{\varphi\varphi})$ data necessary for the AEP~\eqref{ep} is obtained from the 3+1 data in~\eqref{em}, $(N,\bm \beta,\bm\gamma)$ as follows (see Appendix A for details).
 $1)$  $g_{tt} = -N^2 + \gamma_{ij}\beta^i\beta^j$; 
$2)$ $g_{\varphi\varphi}$ determines the perimeter $\mathcal{P}$ of a circumference $r=$constant, 
$\mathcal{P} =2\pi\sqrt{g_{\varphi\varphi}(r)} =\int_0^{2\pi}\sqrt{\gamma_{xx}y^2 - 2\gamma_{xy}xy + \gamma_{yy}x^2}\,d\phi$,
where the auxiliary coordinate $\phi$ is defined via $x=r\cos\phi, y=r\sin\phi$, and may differ from  $\varphi$.
Computing $\mathcal{P}$ numerically via this integral, yields $g_{\varphi\varphi}$; 3)
since $g_{t\varphi}=g_{\varphi\varphi}\beta^\varphi$, then $
g_{t\varphi}= \pm \sqrt{g_{\varphi\varphi}}\,\sqrt{ \gamma_{ij}\beta^i\beta^j}$.
The sign is positive (negative) if $\beta^y>0$ ($\beta^y<0$) when $y=0$.

Finally, two averaging procedures are used: $(1)$ to wash away (mild) axial-symmetry deviations, we work with averaged functions over surfaces with $r=$constant, $e.g.$ 
$
\left<{N^2}\right>\equiv\frac{1}{2\pi}\int_0^{2\pi}N^2\,d\phi.
$
Then, 
$
\left<g_{t\varphi}\right>\equiv \pm \sqrt{g_{\varphi\varphi}}\,\sqrt{\left<{\bm \beta}\cdot{\bm \beta}\right>}$,
and
$
\left<g_{tt}\right>\equiv  -\left<{N^2}\right> + \left<{\bm \beta}\cdot{\bm \beta}\right>
$.
$(2)$ to wash away oscillations in the UCO's time evolution around a trend, we introduce time-averaged potentials $V^*_\pm$:
$
V^*_\pm(r,t)  = [2\,T(t)]^{-1}\int_{t-2\,T(t)}^t V_\pm(r,\tau)\,d\tau$, 
where $T(t)$ is a measure of the oscillation period of the lapse $N(t)$, taken to represent physical oscillations of the evolving UCO.

\noindent{\bf {\em Testing UCO dynamics with bosonic stars (BSs).}} 
Testing the dynamical viability of UCOs, requires models obeying conditions $(i)$ and $(iii)$  in the \textit{Introduction}, to then assess the impact of LRs on condition $(ii)$. The former are obeyed in families of \textit{bosonic stars} (BSs)~\cite{Schunck:2003kk,Brito:2015pxa}, which can form via gravitational cooling~\cite{seidel1994formation,di2018dynamical} and emerge in simple and robust scalar/vector effective field theories minimally coupled to Einstein's gravity~\footnote{The non-self interacting vector model used here is free of known pathologies~\cite{Clough:2022ygm,Coates:2022qia,Mou:2022hqb}.}, where numerical evolutions are under control~\cite{Liebling:2012fv}. Additionally, our analysis requires UCO models free of other instabilities ($e.g.$ perturbative or from ergo-regions), which would mask the impact of LRs. This rules out the simplest, spherical scalar/vector BSs, which only become UCOs in a perturbatively unstable regime~\cite{Cunha:2017wao}.

\begin{figure}[t!]
\centering
\includegraphics[width=1.0\linewidth]{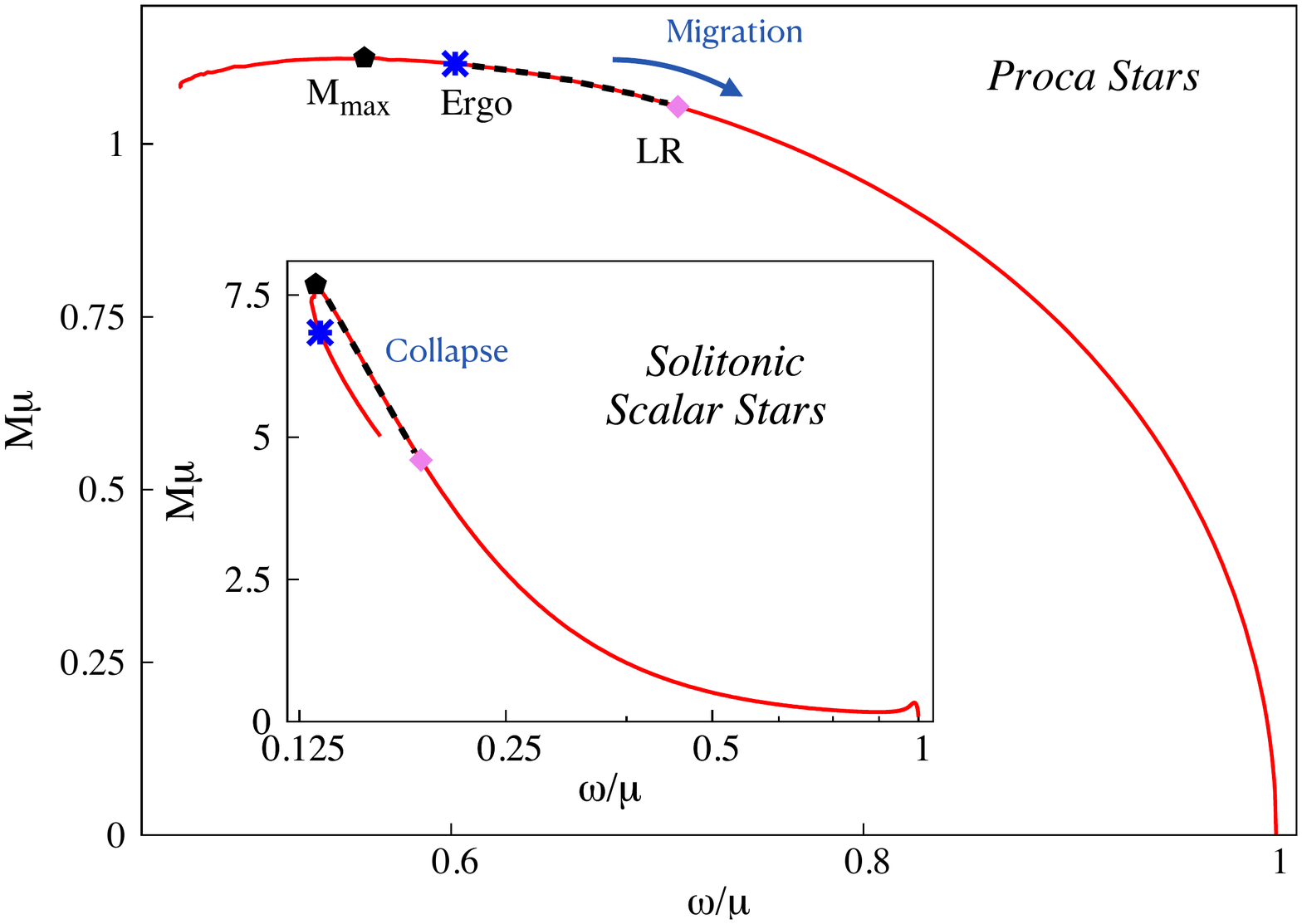}
\caption{Space of solutions for Model 1 (main panel) and Model 2 (inset). Starting at $\omega/\mu=1$ and moving leftwards along the solutions curves, one encounters the first solution with a LR (ergo-region) at the pink diamond (blue star). The maximal mass BS, at which an unstable mode is expected to appear (see $e.g.$~\cite{Siemonsen:2020hcg}) is marked with a black pentagon.}
\label{figBSs}
\end{figure}

We have identified two appropriate models of BSs for studying the LR instability, both described by the Lagrangian density,  $\mathcal{L}=R/(16\pi G)+\mathcal{L}_m$, where $R$, $G$ are the Ricci scalar and Newton's constant~\footnote{We use geometrized units, $G=1=c$ here after.}. 
Model 1 has $\mathcal{L}_m=-\mathcal{F}_{\alpha\beta}\bar{\mathcal{F}}^{\alpha\beta}/4-\mu^2\mathcal{A}_{\alpha}\bar{\mathcal{A}}^{\alpha}/2$, describes a complex Proca model with mass $\mu$~\footnote{Overbar denotes complex conjugation.} and we focus on its fundamental, spinning, mini-Proca star solutions~\cite{Herdeiro:2017phl,Herdeiro:2019mbz}. They are labelled by their ADM mass $M$ or their oscillation frequency $\omega$ and are dynamically robust~\cite{sanchis2019nonlinear}. These Proca stars become UCOs for $\omega/\mu \lesssim 0.711$; an ergoregion emerges for $\omega/\mu \lesssim 0.602$ and perturbative instabilities are expected beyond the maximal $M$, for $\omega/\mu \lesssim 0.562$ - Fig.~\ref{figBSs} (main panel). 
Model 2 has $\mathcal{L}_m=  -\partial_\alpha \Phi  \partial^\alpha\bar{\Phi}  - \mu^2 |\Phi|^2\left[1- 2 |\Phi|^2/\sigma_0^2 \right]^2$, describes a self-interacting complex scalar model with mass $\mu$ and coupling $\sigma_0$. We focus on its fundamental, spinning, solitonic boson star solutions~\cite{Kleihaus:2005me,Siemonsen:2020hcg}, again labelled by $M$ or $\omega$, choosing the illustrative value $\sigma_0= 0.05$~ \cite{Bezares:2017mzk}. There is a  dynamically robust relativistic branch of solutions for $\omega/\mu \lesssim 0.493$~\cite{Siemonsen:2020hcg}, that become UCOs for $\omega/\mu \lesssim 0.188$;  perturbative instabilities are expected beyond the maximal M, for $\omega/\mu \lesssim 0.132$ and the ergo-region appears in this unstable branch for $\omega/\mu \lesssim 0.134$  - Fig.~\ref{figBSs} (inset)~\footnote{In both models solutions with ergo-regions are UCOs, in agreement with~\cite{Ghosh:2021txu}.}. 

The just described Proca (solitonic boson) stars in the range $0.602<\omega/\mu<0.711$  ($0.132<\omega/\mu<0.188$) provide tests for the LR instability (dashed line segments in Fig.~\ref{figBSs}). Using them as initial data, we have performed fully non-linear numerical evolutions of the Einstein-bosonic systems with spacetime variables in the BSSN formulation and the \textsc{Einstein Toolkit}~\cite{EinsteinToolkit:web,loffler2012f}  evolved using
McLachlan~\cite{Brown:2008sb,McLachlan:web} and \textsc{Lean}~\cite{ZilhaoWitekCanudaRepository} - codes available in~\cite{Canuda_zenodo_3565474} and described in~\cite{Zilhao:2015tya,Cunha:2017wao,sanchis2019head}.

\begin{figure}[h!]
\centering
\includegraphics[width=1.0\linewidth]{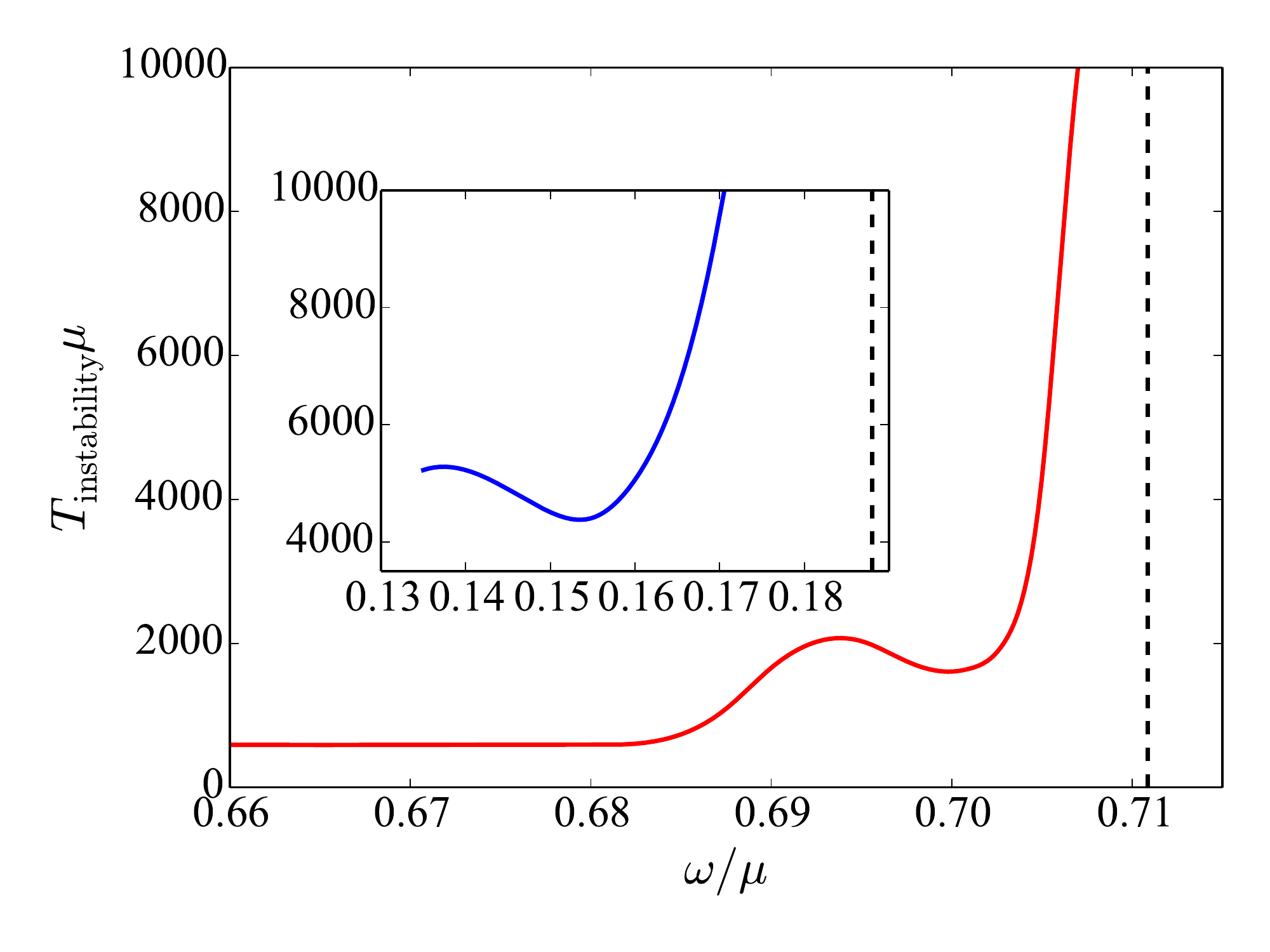}
\caption{Instability timescale for Model 1 (main panel) and Model 2 (inset). The dashed vertical line marks the first UCO.}
\label{figTimeinstability}
\end{figure}

{\bf {\em Results Model 1: migration.}} 
We have evolved the aforementioned Proca stars up to $t\mu=10^4$ and confirmed the existence of an instability for the UCOs. Fig.~\ref{figTimeinstability} (main panel) shows the time at which the instability starts $vs.$ $\omega$. The timescale tends to diverge when approaching the first star with a LR ($\omega/\mu\simeq 0.711$), thus associating the instability with UCOs.  For some of the models with LR we could not see the development of instability, since the simulations last only up to $t\mu=10^4$ but our results suggest that they will become unstable if evolved for longer. The instability is not seen for stars with $\omega/\mu> 0.711$.

\begin{figure}[t!]
\begin{tabular}{ p{0.41\linewidth} p{0.30\linewidth}  p{0.21\linewidth} }
\centering  
$\ \ \ \  \ \omega/\mu=0.68$&\centering $\ \ \ \ \omega/\mu=0.69$ &\centering $\omega/\mu=0.70$
\end{tabular}
\centering
\includegraphics[width=0.245\linewidth]{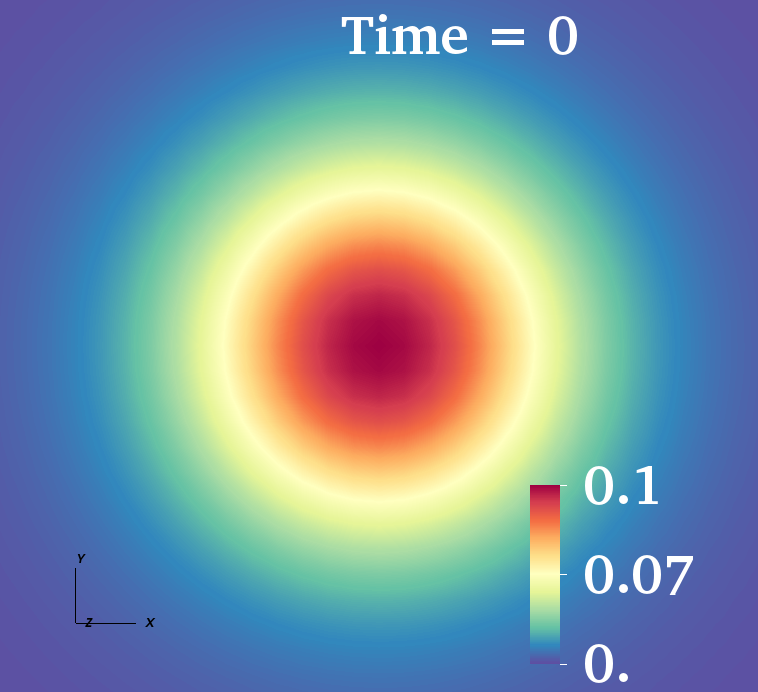}\hspace{-0.01\linewidth}
\includegraphics[width=0.245\linewidth]{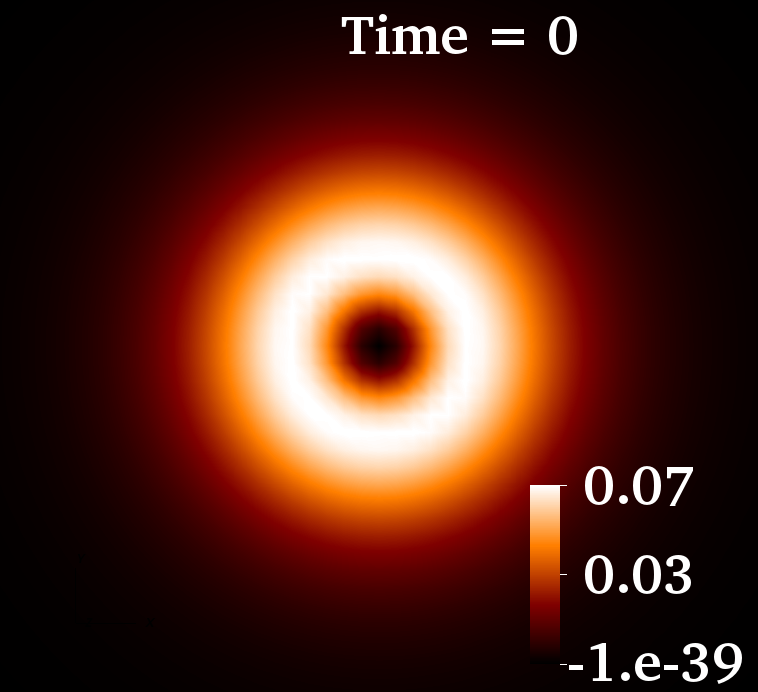}\hspace{-0.01\linewidth} \
\includegraphics[width=0.245\linewidth]{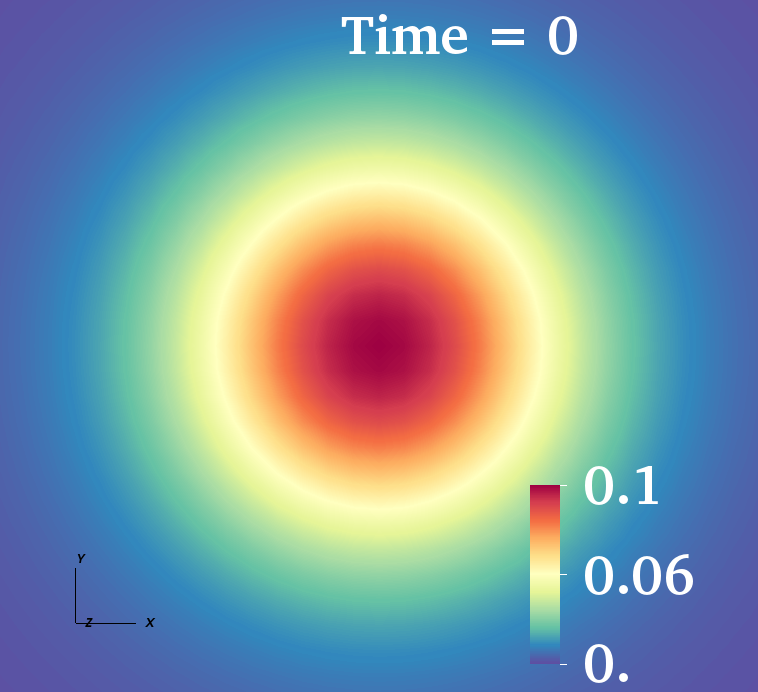}\hspace{-0.01\linewidth}
\includegraphics[width=0.245\linewidth]{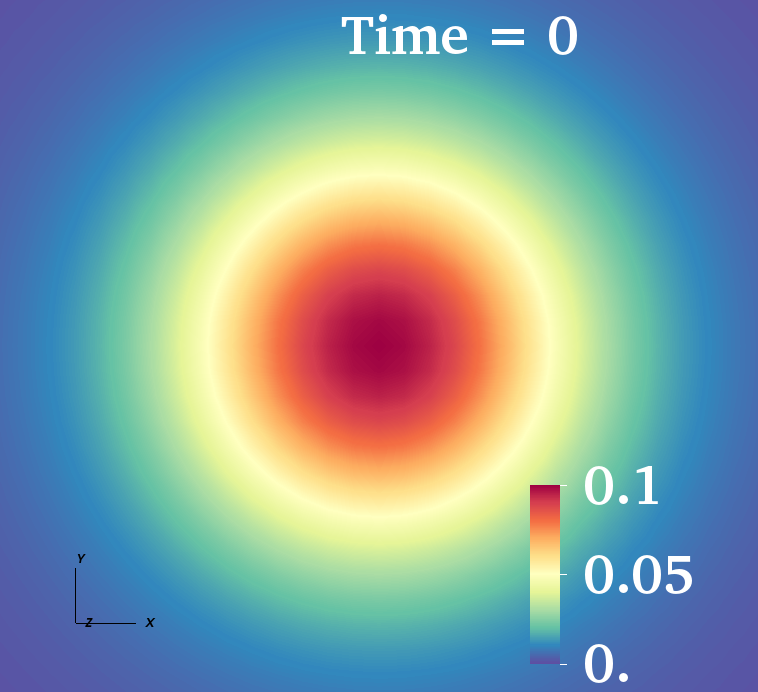}\hspace{-0.01\linewidth}\\
\includegraphics[width=0.245\linewidth]{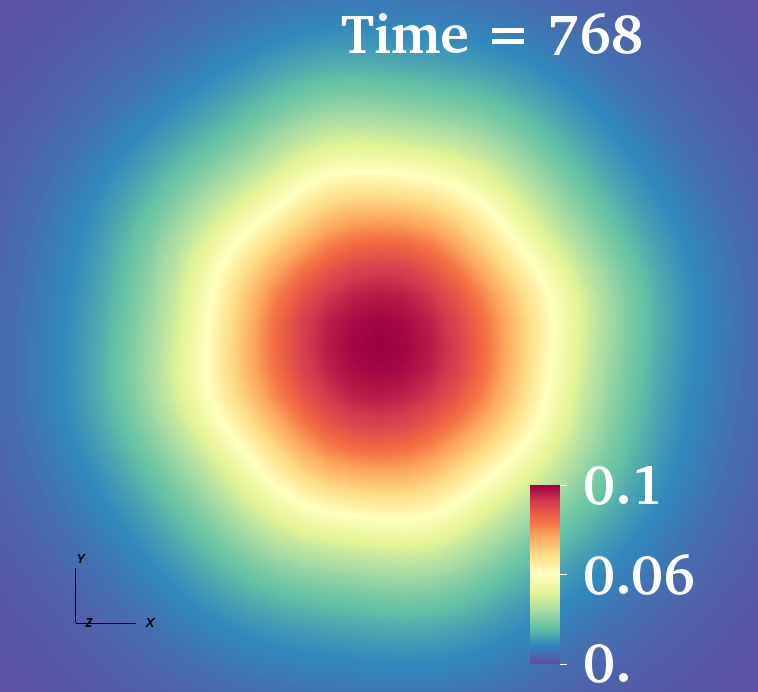}\hspace{-0.01\linewidth}
\includegraphics[width=0.245\linewidth]{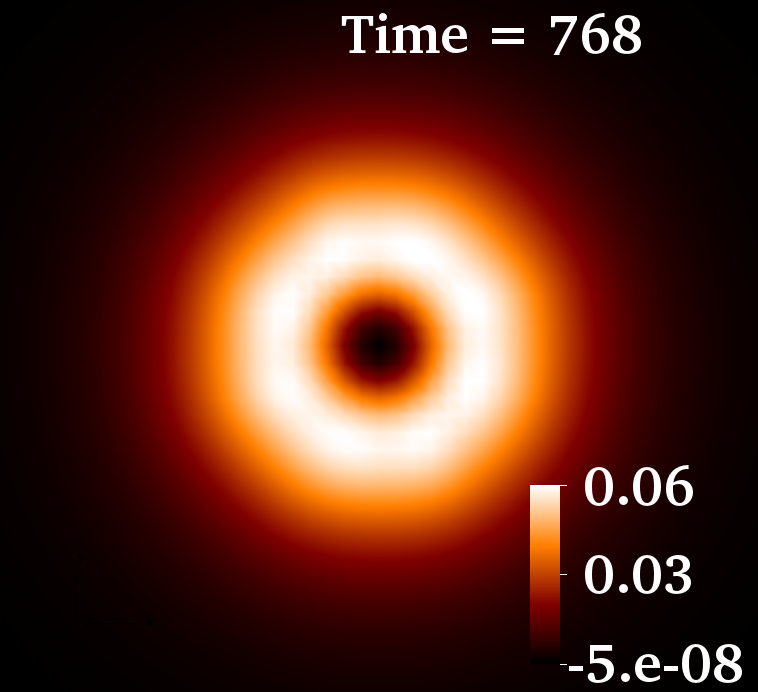}\hspace{-0.01\linewidth} \
\includegraphics[width=0.245\linewidth]{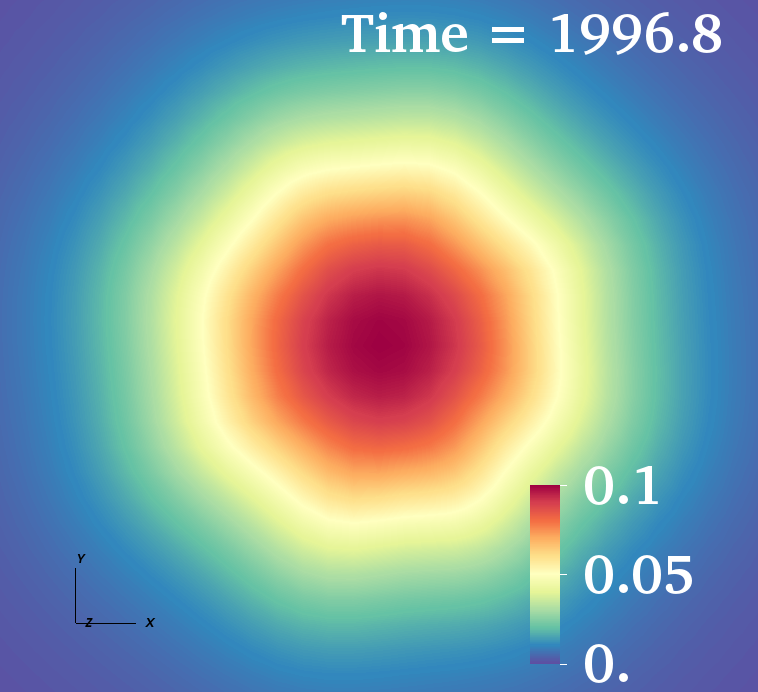}\hspace{-0.01\linewidth}
\includegraphics[width=0.245\linewidth]{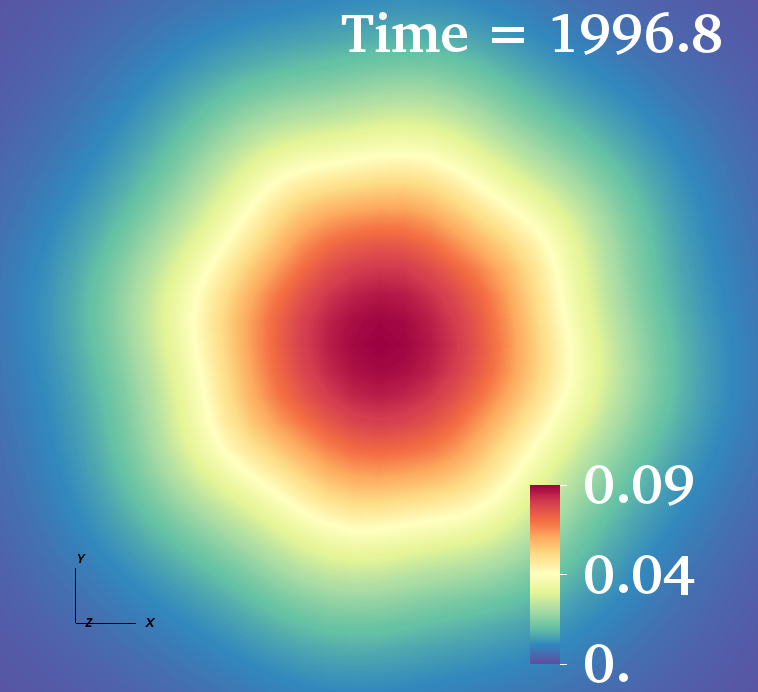}\hspace{-0.01\linewidth}\\
\includegraphics[width=0.245\linewidth]{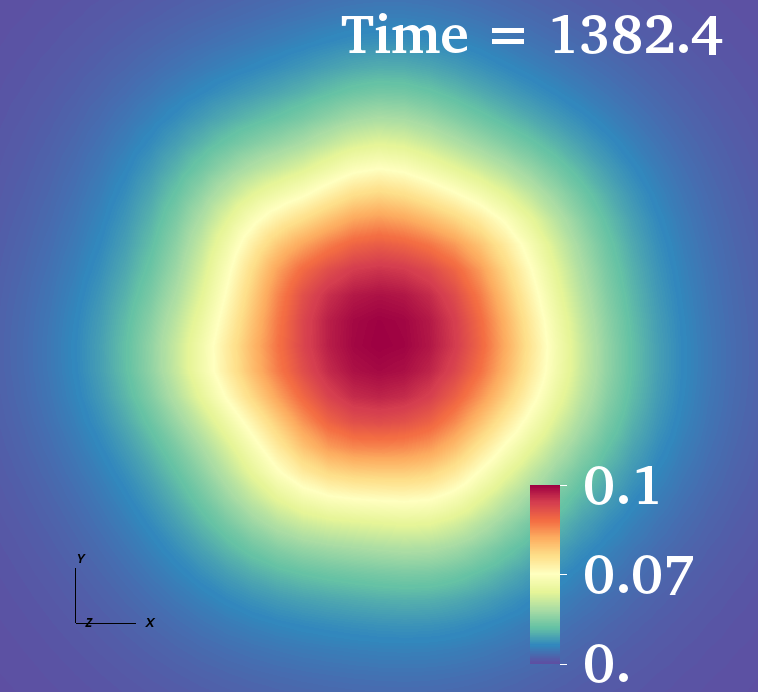}\hspace{-0.01\linewidth}
\includegraphics[width=0.245\linewidth]{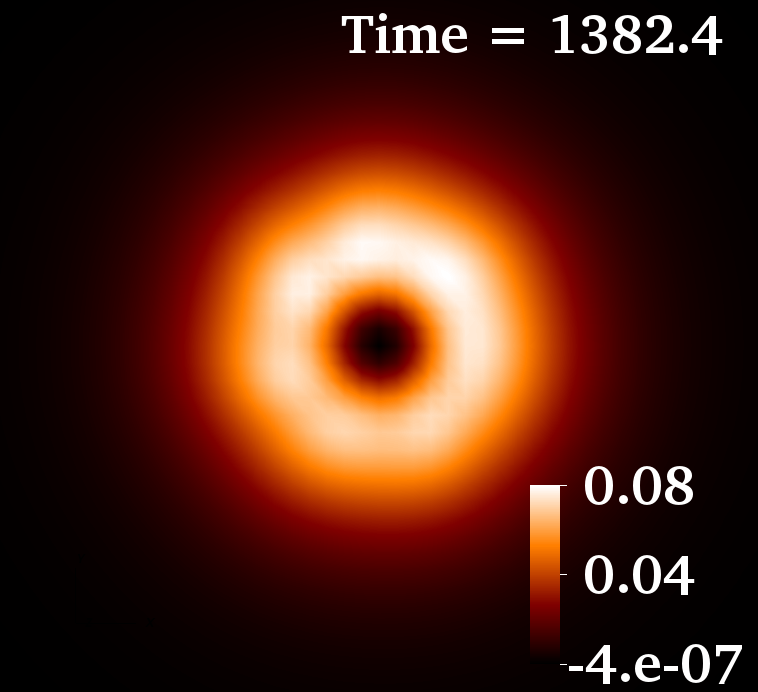}\hspace{-0.01\linewidth} \
\includegraphics[width=0.245\linewidth]{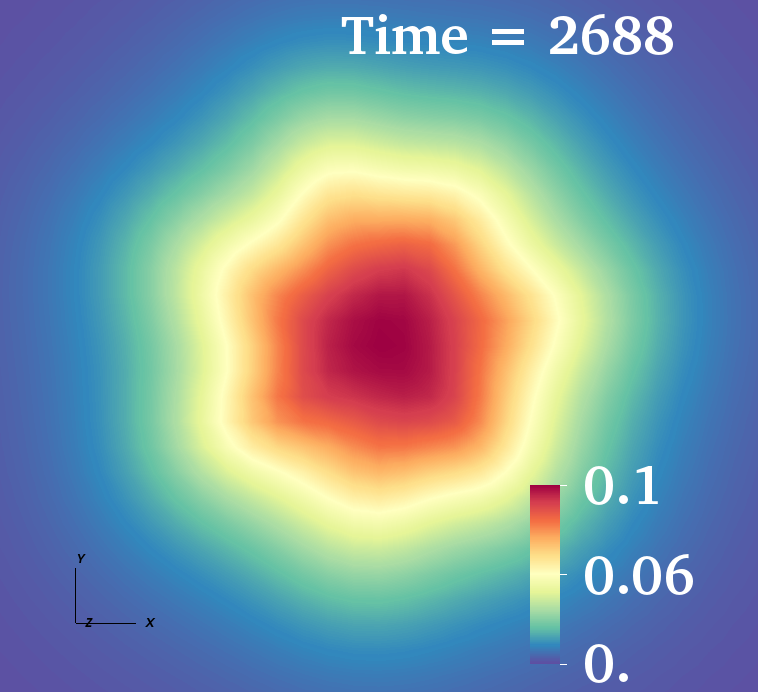}\hspace{-0.01\linewidth}
\includegraphics[width=0.245\linewidth]{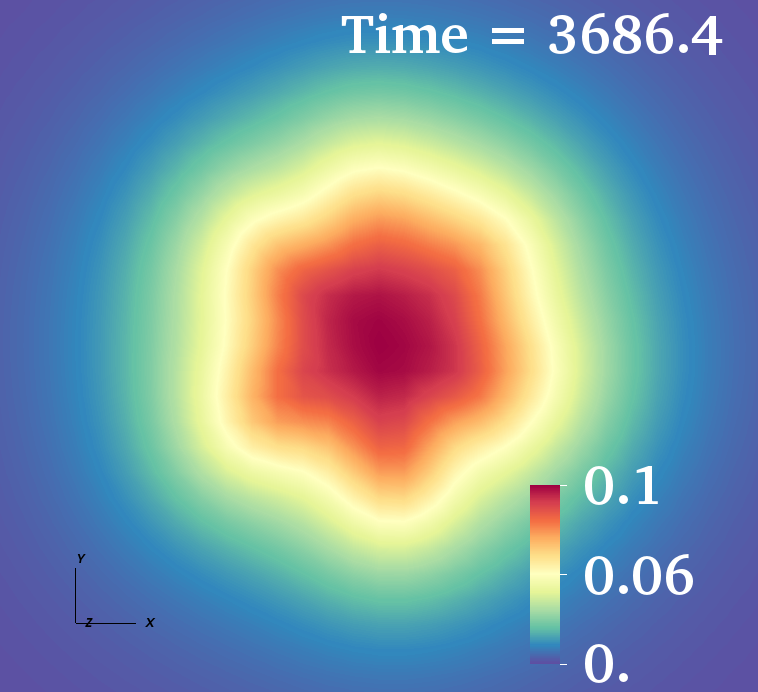}\hspace{-0.01\linewidth}\\
\includegraphics[width=0.245\linewidth]{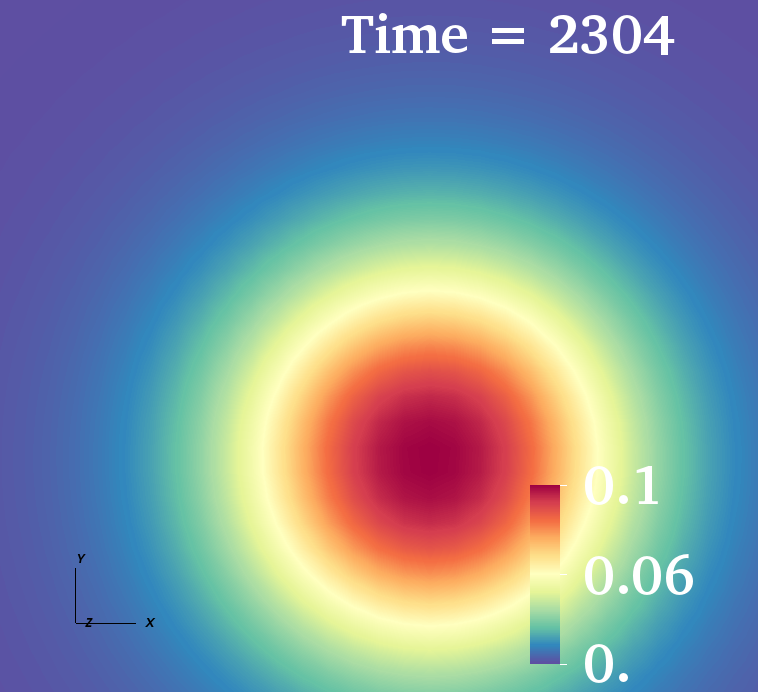}\hspace{-0.01\linewidth}
\includegraphics[width=0.245\linewidth]{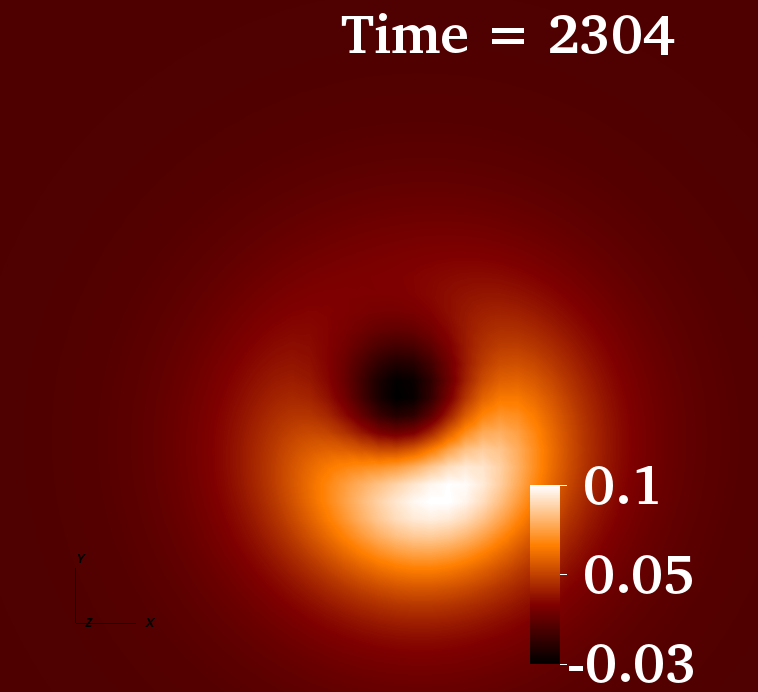}\hspace{-0.01\linewidth} \ 
\includegraphics[width=0.245\linewidth]{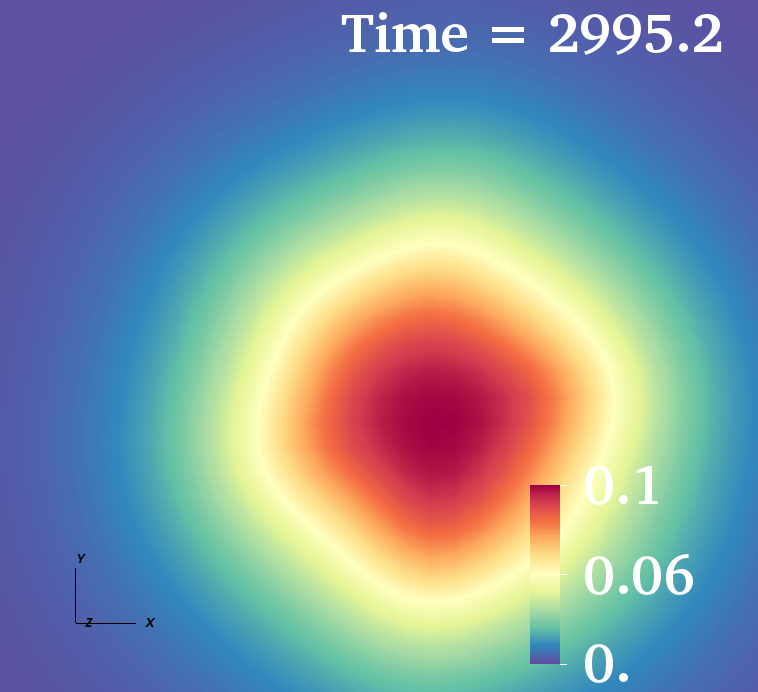}\hspace{-0.01\linewidth}
\includegraphics[width=0.245\linewidth]{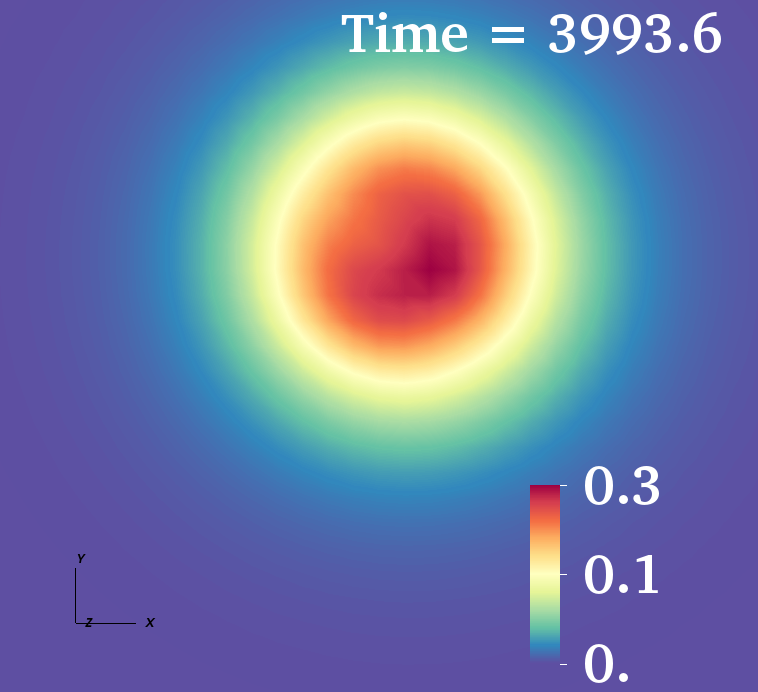}\hspace{-0.01\linewidth}\\
\includegraphics[width=0.245\linewidth]{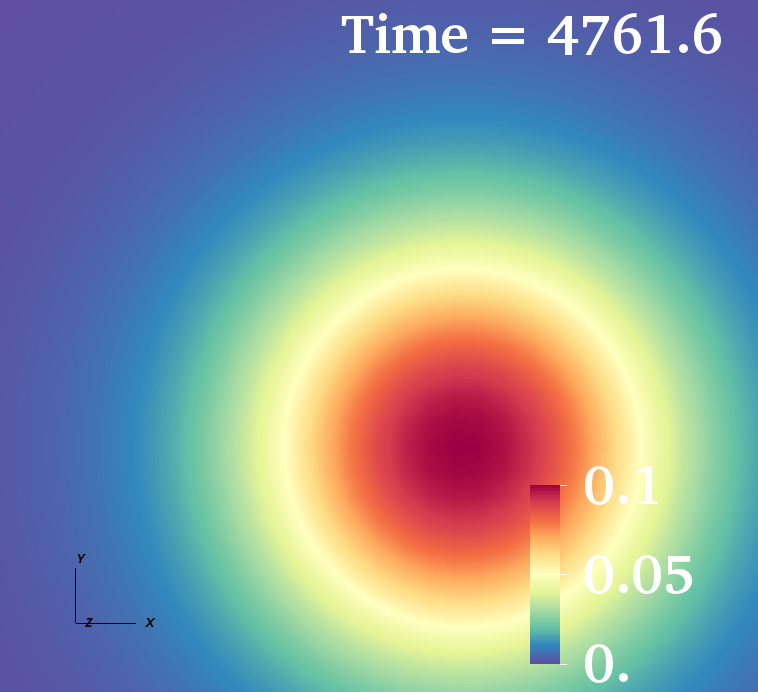}\hspace{-0.01\linewidth}
\includegraphics[width=0.245\linewidth]{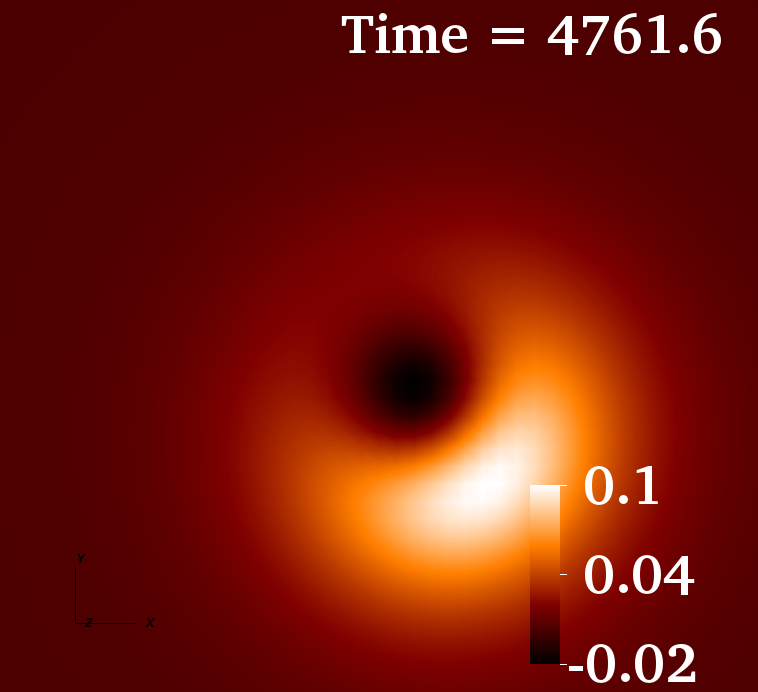}\hspace{-0.01\linewidth} \ 
\includegraphics[width=0.245\linewidth]{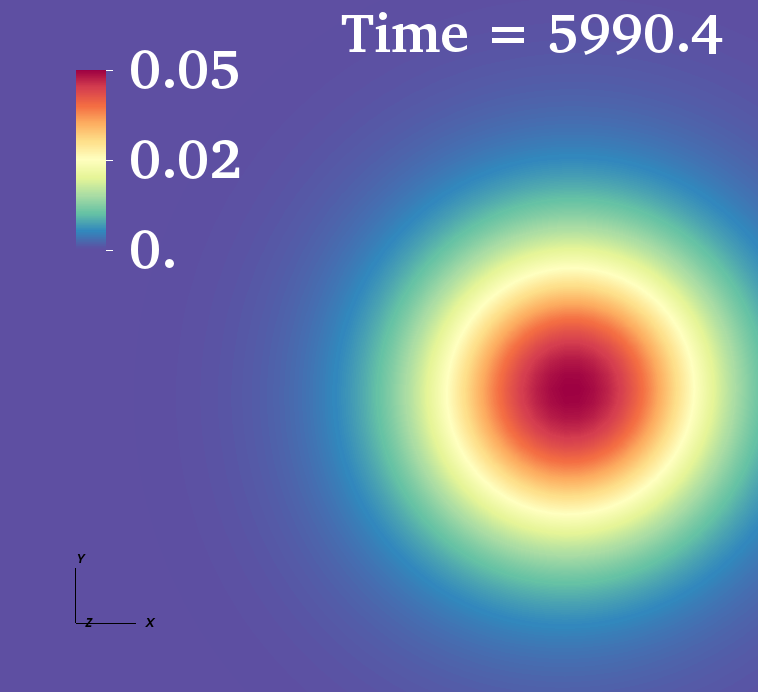}\hspace{-0.01\linewidth}
\includegraphics[width=0.245\linewidth]{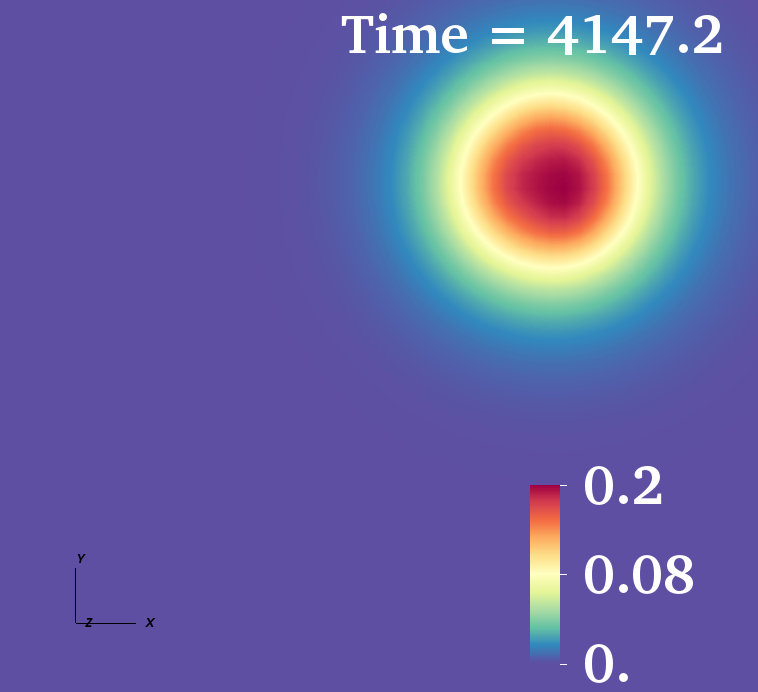}\hspace{-0.0\linewidth}
\caption{Snapshots of the energy density (and angular momentum density in middle left column) of three spinning Proca Star UCOs with $\omega/\mu=0.68$ (left and midlle left column), 0.69 (middle right column) and 0.70 (right column). Time runs from top to bottom, given in code units.}
\label{figProcamodels}
\end{figure}

To grasp the nature of the instability, Fig.~\ref{figProcamodels} shows snapshots of the time evolution of the energy density  on the equatorial plane for stars with  $\omega/\mu=0.68$, 0.69, 0.70, and angular momentum density for $\omega/\mu=0.68$. The energy density first acquires an octogonal shape (see second row of Fig.~\ref{figProcamodels}) which evolves into a ``starfish-like" pattern (third row). The axisymmetry of the star is mildly broken and the star suffers a small kick, slowly moving away from the grid center. The instability triggers a balanced mass and angular momentum loss, mostly carried by the Proca field, with almost no gravitational wave emission.
This balanced loss allows a migration to less massive and compact but still spinning Proca stars without LRs - see arrow in Fig.~\ref{figBSs} and Appendix B for details.

\begin{figure}[h!]
\begin{center}
\includegraphics[width=0.5\textwidth]{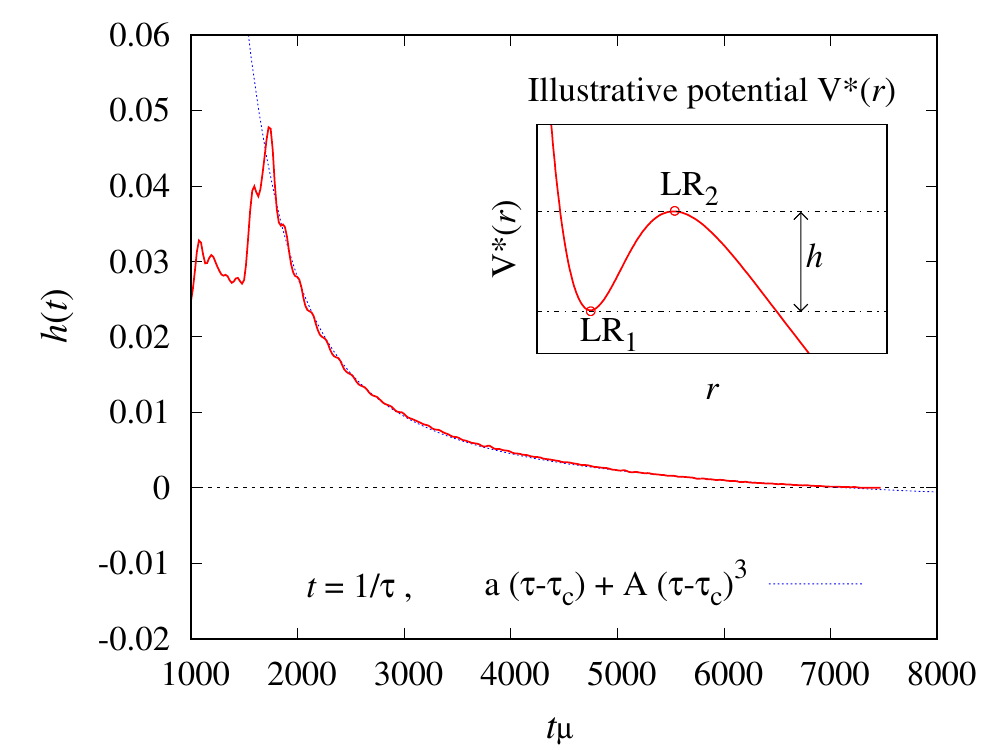}
\includegraphics[width=0.5\textwidth]{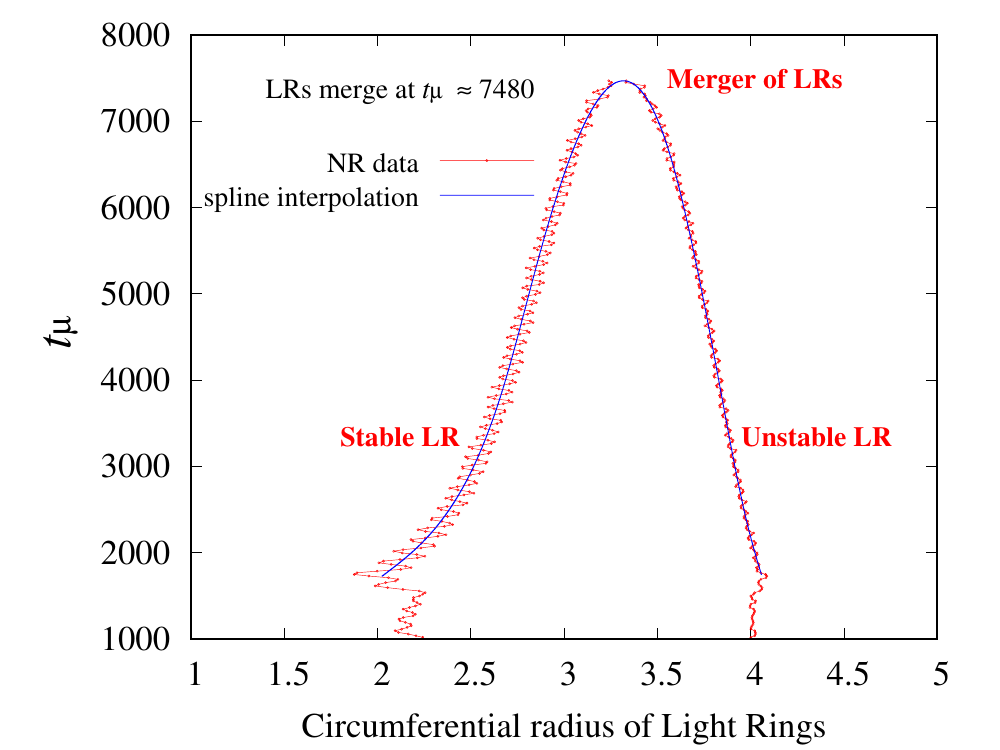}
\caption{\small (Top) Potential-well depth $h$ (illustrated in inset) as a function of time for $\omega/\mu=0.68$. A guiding fit function (dotted line) is defined in terms of the inverse time $\tau=1/t$, with adjusted constants $\{a,A,\tau_c\}$. (Bottom) The LRs' circumferential radii $\sqrt{g_{\varphi\varphi}}/M_t$  for different values of time $t$. The blue line is an auxiliary spline interpolation to help convey the underlying pattern.
The time-averaged potential $V_-^*$ can suppress several, but not all, oscillation modes of the raw potential $V_-$. This is the likely cause of the observed zig-zagging of the red curve, a manifestation of residual star radial oscillations around their unexcited state (at that time).}
\label{depth-plot}
\end{center}
\end{figure}

Further clear evidence that the fate of this instability is a non-UCO spinning Proca star is obtained by examining the averaged AEP $V^*_-$~\footnote{For these Proca stars the LRs emerge as critical points of $V_-$, rather than $V_+$.}. At each time slice, this potential has the typical radial profile for an UCO displaying a stable and an unstable LR - Fig.~\ref{depth-plot}  (top - inset) - from which one can extract a potential depth $h$, defined as the (positive) potential difference of $V^*_-$ computed at the LRs. $h$ can be represented as a function of time in the evolutions. Fig.~\ref{depth-plot}  (top) shows it for $\omega/\mu=0.68$. It is clear that the depth $h(t)$ reaches zero at $t_c\,\mu \simeq 7480$, $i.e.$ the two LRs have merged in a finite time. This merger is further seen on the bottom panel of Fig.~\ref{depth-plot}, where the perimetral radius of both LRs converges to a single value. After this point the LRs essentially disappear in terms of $V^*_-$. However, for a limited couple of instances right after $t>t_c$ the LRs can be recreated and destroyed again due to sporadic fluctuations of the potential (not shown in Fig.~\ref{depth-plot}), before disappearing altogether. This confirms the migration robustly evolves in the direction of destroying the LR pair. In Appendix A, the time evolution of the AEP is further detailed, legitimating the validity of the adiabatic approximation underlying the AEP.

%

\begin{figure}
\centering
\includegraphics[width=0.245\linewidth]{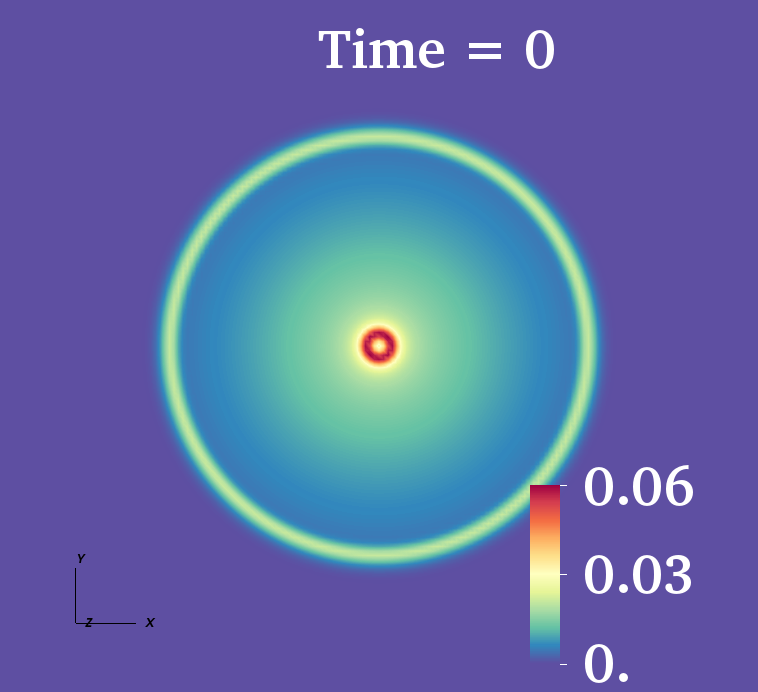}\hspace{-0.01\linewidth}
\includegraphics[width=0.245\linewidth]{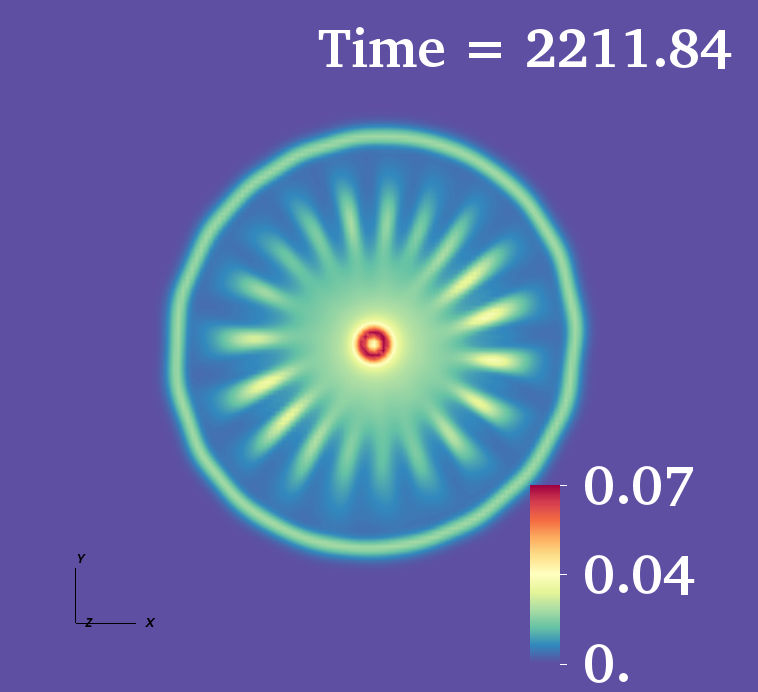}\hspace{-0.01\linewidth} \
\includegraphics[width=0.245\linewidth]{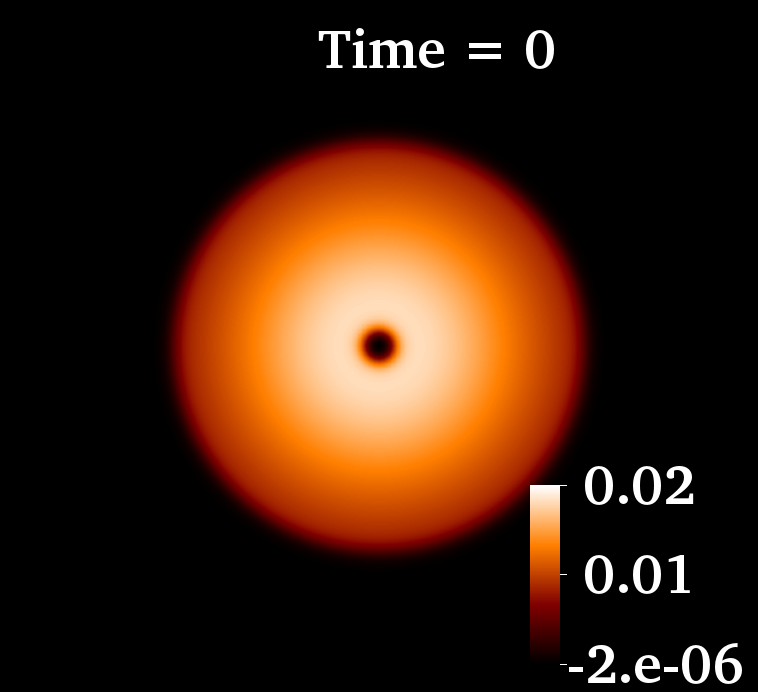}\hspace{-0.01\linewidth}
\includegraphics[width=0.245\linewidth]{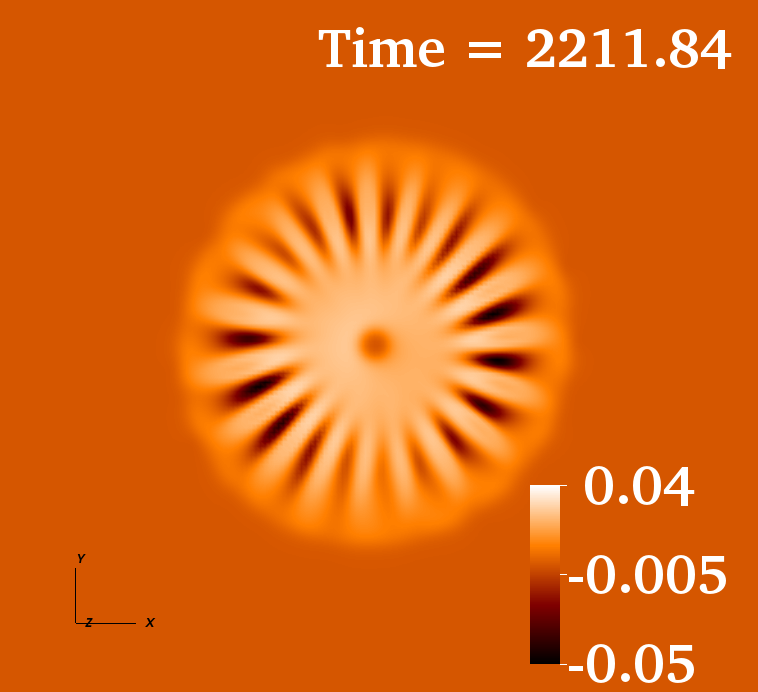}\hspace{-0.01\linewidth}\\
\includegraphics[width=0.245\linewidth]{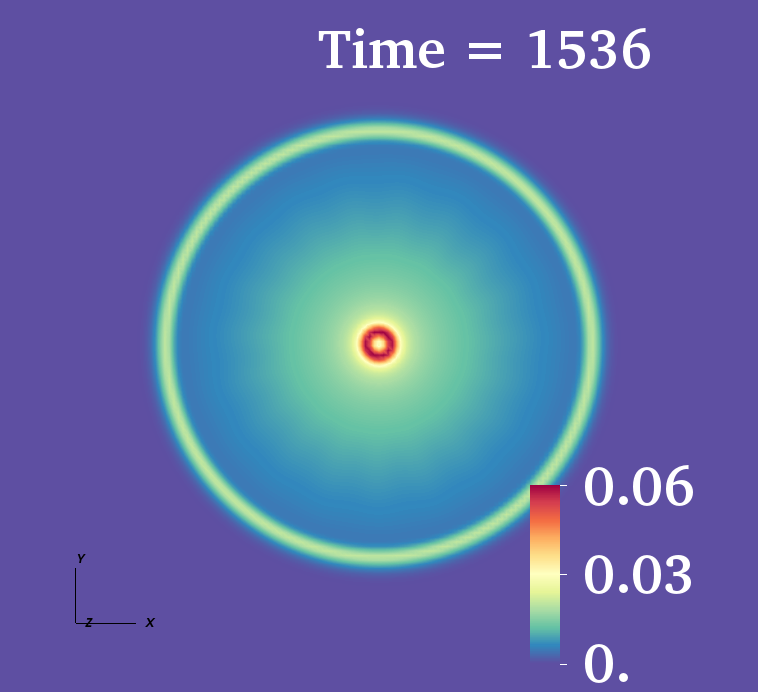}\hspace{-0.01\linewidth}
\includegraphics[width=0.245\linewidth]{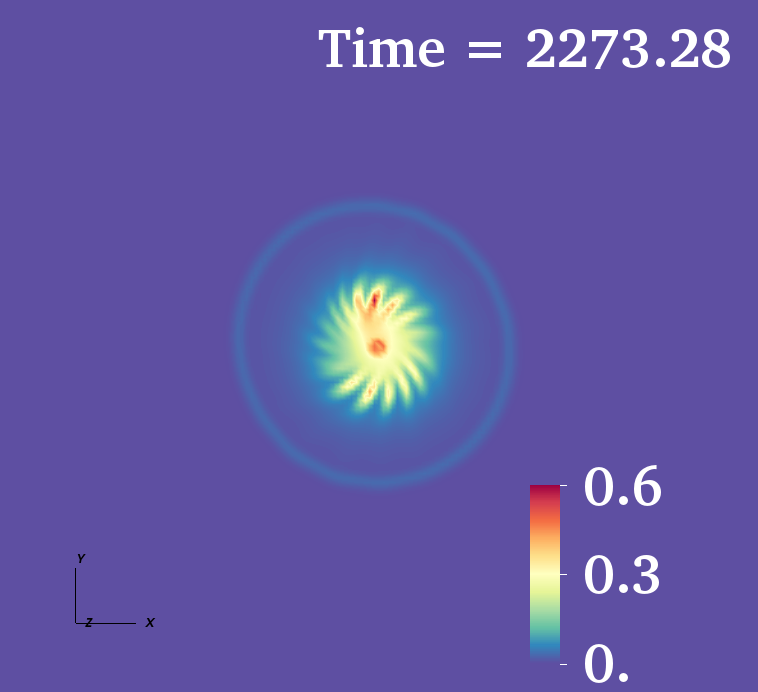}\hspace{-0.01\linewidth} \
\includegraphics[width=0.245\linewidth]{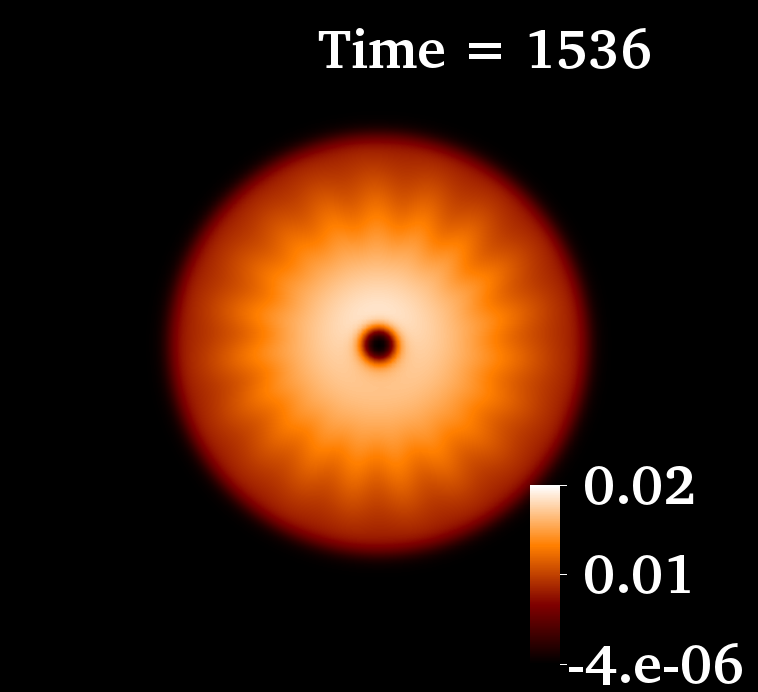}\hspace{-0.01\linewidth}
\includegraphics[width=0.245\linewidth]{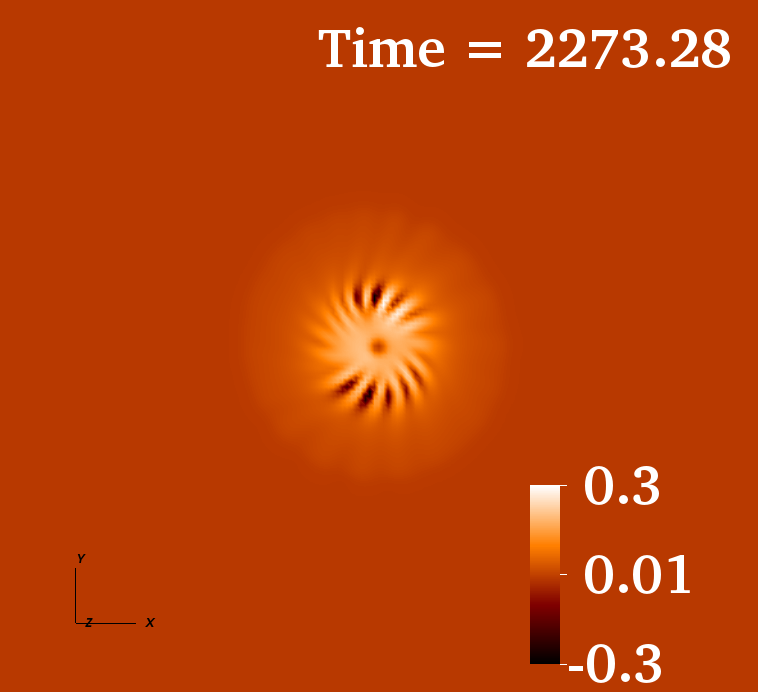}\hspace{-0.01\linewidth}\\
\includegraphics[width=0.245\linewidth]{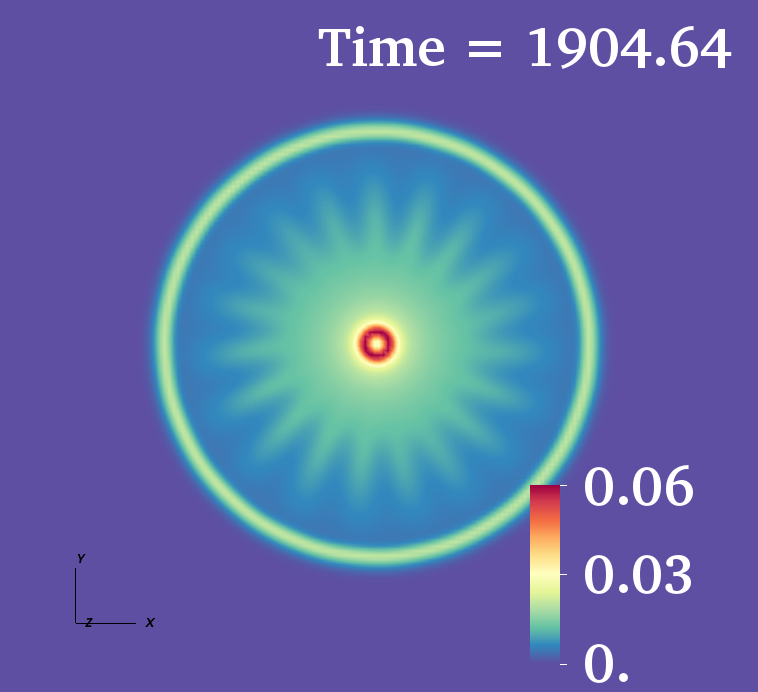}\hspace{-0.01\linewidth}
\includegraphics[width=0.245\linewidth]{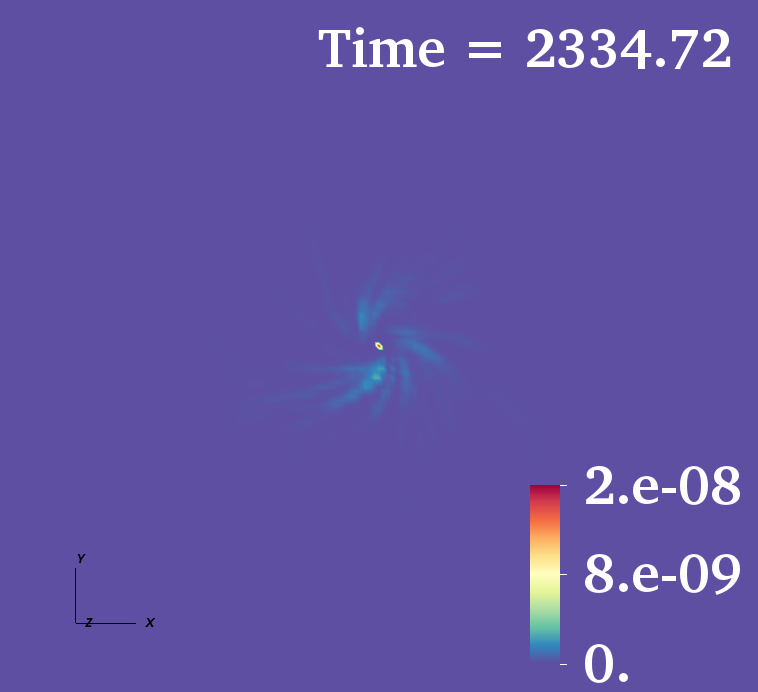}\hspace{-0.01\linewidth} \
\includegraphics[width=0.245\linewidth]{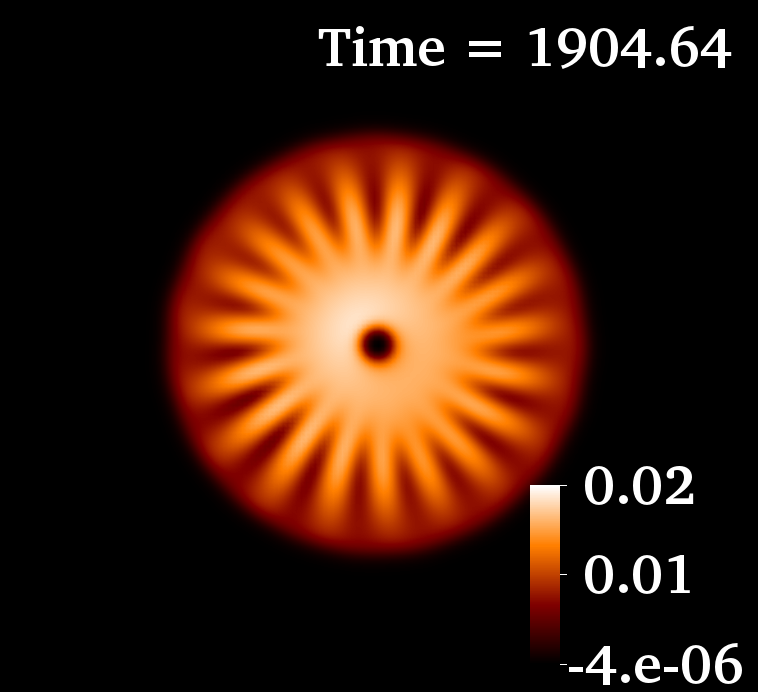}\hspace{-0.01\linewidth}
\includegraphics[width=0.245\linewidth]{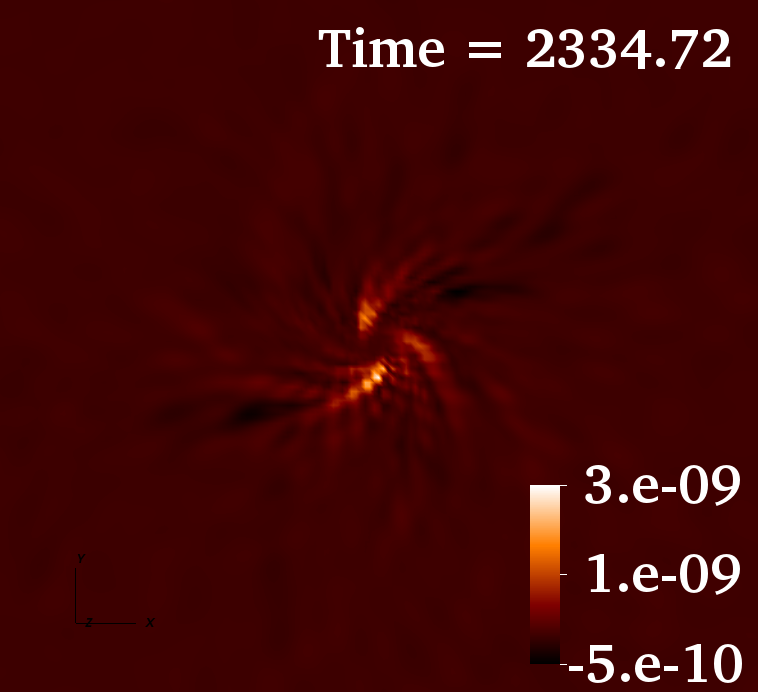}\hspace{-0.01\linewidth}\\
\caption{Time evolution of a spinning solitonic BSs with  $\omega/\mu=0.16$ which has LRs. (Left and middle left column) Energy density. Time runs from top to bottom and then from left to right. (Middle right and right column) Angular momentum density. Both quantities are extracted on the equatorial plane.}
\label{figScalar}
\end{figure}

{\bf {\em Results Model 2: collapse.}} 
We have also evolved the spinning solitonic scalar boson stars described above. In the UCO region,  $\omega/\mu\lesssim 0.188$, we have again observed an instability, absent for $\omega> 0.188$ (but within the relativistic stable branch). Its timescale again tends to diverge, as we approach the critical frequency - Fig.~\ref{figTimeinstability} (inset), correlating the instability with the existence of LRs.  The initial development of the instability qualitatively resembles the previous case in the loss of axi-symmetry - see Fig.~\ref{figScalar} for evolution snapshots of the star with $\omega/\mu=0.16$. However, the outcome is different; the star collapses and forms a spinning BH, diagnosed by the emergence of an apparent horizon (detailed in Appendix B). The culprit of this different outcome may be the self-interaction potential, that confines the star and suppresses the dissipation through gravitational cooling, the essential channel by which migration occurs in model 1. The AEP shows violent oscillations and appears to deepen near the location of the stable LR before the collapse, implying the technique looses validity.

{\bf {\em Remarks.}} 
Proving the BH hypothesis is as challenging as disproving the ECO hypothesis. A smaller, but informative step, is to rule out classes of inadequate models as BH alternatives. 
The evidence shown in this letter supports the inadequacy, as BH foils, of a large class of UCOs, for which a plausible formation mechanism exists via an incomplete gravitational collapse of quasi-Minkowski initial data. Even if such UCOs could form as a transient state, their unavoidable stable LR triggers an instability   that, generically, develops in a moderate time scale, either promoting collapse to BHs or migration to a non-UCO. It remains to be seen whether UCOs near the critical configuration, for which the timescale can grow (likely) arbitrarily large, still retain any effectiveness as BH foils, and if other fates are possible for the LR instability, $e.g.$ considering different UCO models. 
It also remains an open problem to find a practical estimate of the LR instability timescale in a generic spacetime. One of our main results is a concrete timescale within two specific models. The difficulty  arises from the putative non-linear character of the instability. Perhaps some  lesson can be taken from other gravitational non-linear instabilities, namely the Anti-de Sitter (AdS) turbulent instability~\cite{Bizon:2011gg}. In fact, the trapping of modes in the stable LR potential well has some conceptual similarities with the trapping of perturbations in the AdS ``box".

There are mechanisms to avoid the conclusion in~\cite{Cunha:2017qtt} that UCOs have stable LRs, $e.g.$: $(i)$ topological non-triviality; $e.g.$ ultracompact wormholes, since such UCOs are not diffeomorphic to Minkowski spacetime. But any mechanism forming such macroscopic UCOs is beyond General Relativity and remains to be established; $(ii)$ non-axisymmetry; but the known astrophysical compact objects exhibit to a high degree axisymmetry; $(iii)$ non-circularity; but very few (analytic or numerical) solutions describing compact objects in Einstein's or modified gravity without circularity are known.

The ability of ECOs to imitate BH phenomenology does not \textit{demand} LRs. Examples are known, both for gravitational wave ($e.g.$~\cite{CalderonBustillo:2020srq,CalderonBustillo:2022cja}) and electromagnetic ($e.g.$~\cite{Herdeiro:2021lwl}) observations, where ECOs without LRs could mimic, in limited circumstances, some strong gravity BH features. It seems implausible, however, that ECOs without LRs may mimic \textit{all} BH phenomenology and replace them entirely as physical players in the Universe. Under this rationale, our results challenge the ECO hypothesis as a complete replacement of the BH hypothesis, but not as another ingredient of the physical Universe.


\bigskip

{\bf {\em Acknowledgements.}} 
This work was supported by the Spanish Agencia Estatal de Investigaci\'on (grant PGC2018-095984-B-I00),
by the Generalitat Valenciana (PROMETEO/2019/071), by the Center for Research and Development in Mathematics and Applications (CIDMA) through the Portuguese Foundation for Science and Technology (FCT - Funda\c c\~ao para a Ci\^encia e a Tecnologia), references UIDB/04106/2020 and UIDP/04106/2020, by national funds (OE), through FCT, I.P., in the scope of the framework contract foreseen in the numbers 4, 5 and 6
of the article 23, of the Decree-Law 57/2016, of August 29,
changed by Law 57/2017, of July 19 and by the projects PTDC/FIS-OUT/28407/2017,  CERN/FIS-PAR/0027/2019, PTDC/FIS-AST/3041/2020 and CERN/FIS-
PAR/0024/2021. This work has further been supported by  the  European  Union's  Horizon  2020  research  and  innovation  (RISE) programme H2020-MSCA-RISE-2017 Grant No.~FunFiCO-777740 and by FCT through
Project~No.~UIDB/00099/2020.
PC is supported by the Individual CEEC program 2020 funded by the FCT.
N.S.G. acknowledges financial support by the Spanish Ministerio de Universidades, reference UP2021-044, within the European Union-Next Generation EU.
Computations have been performed at the Servei d'Inform\`atica de la Universitat
de Val\`encia, and the Argus and Blafis cluster at the U. Aveiro.

\bigskip

{\bf {\em Appendix A. Details on the Adiabatic Effective Potential (AEP).}} 
We shall consider a 3+1 spacetime metric satisfying the following assumptions:

\textit{[Adiabatic approximation]}
The spacetime at each time slice $t$ is approximately stationary, and thus exists an approximate Killing vector $\partial_t$ connected to the time coordinate $t$. Under this assumption, the evolution is regarded as adiabatic.

\textit{[Axial-symmetry]}
The spacetime is approximately axially-symmetric, and thus exists an approximate Killing vector $\partial_\varphi$ connected to this assumption.

\textit{[Circularity \& Asymptotic flatness]}
The spacetime satisfies approximate circularity and asymptotic flatness. Both these assumptions imply the spacetime admits an (approximate) 2-space orthogonal to $\{\partial_t, \partial_\varphi\}$ and possesses a discrete symmetry $(t, \varphi) \to (-t, -\varphi)$. This also implies that each point exists $\lambda\in\mathbb{R}$ such that $\partial_\varphi = \lambda {\bm\beta}$.

\textit{[Adaptation to axial-symmetry]}
The approximate Killing vector $\partial_\varphi$ is assumed to be tangent to surfaces with $x^2+y^2\equiv r^2= constant$. The coordinate center $\mathcal{O}$: $(x,y)=(0,0)$ is linearly shifted at each time $t$ to the center of the gravitational field distribution, using the lapse $N$ as a reference.

\textit{[Equatorial symmetry]}
The spacetime is assumed to possess a $\mathbb{Z}_2$ reflection symmetry on the equatorial plane surface $z=0$. Together with the assumed axial-symmetry, metric data on this geometrically invariant surface can be directly compared across different times. For this reason, and considering that LRs are also typically found on the equatorial plane, the AEP analysis in this work will be restricted to the surface $z=0$ (Equatorial plane).

Given these assumptions, one can readily note that:
\begin{align}
\label{gttEq}&g_{tt} = -N^2 + \gamma_{ij}\beta^i\beta^j \equiv -N^2 + {\bm \beta}\cdot{\bm \beta}\\
&g_{t\varphi} = \gamma_{\varphi\varphi}\beta^\varphi = g_{\varphi\varphi}\beta^\varphi = \beta_\varphi\\
& g_{r\varphi} = \partial_r\cdot \partial_\varphi = 0
\end{align}
The metric component $g_{tt}$ can be directly obtained from the 3+1 numerical relativity data. In contrast, obtaining both $g_{t\varphi}$ and $g_{\varphi\varphi}$ from the numerical relativity data, necessary to define the AEP, is less straightforward. Critically, one can realise that the perimeter $\mathcal{P}$ of a circumference of constant coordinate $r$ is determined by $g_{\varphi\varphi}$. Writing the 2-metric on the equatorial plane as
\begin{align*}
ds_{(2)}^2&=g_{\varphi\varphi}d\varphi^2 + g_{rr}dr^2 \\
&= \gamma_{xx}dx^2 + 2\gamma_{xy}dxdy + \gamma_{yy}dy^2\,,
\end{align*}
we can obtain the integral of the proper distance over a circle of constant coordinate $r$:
\begin{align*}
\label{Pphi}\mathcal{P} &= \oint_{r=\textrm{const.}}ds=2\pi\sqrt{g_{\varphi\varphi}(r)} \\
&= \int_0^{2\pi}\sqrt{\gamma_{xx}y^2 - 2\gamma_{xy}xy + \gamma_{yy}x^2}\,d\phi\,.
\end{align*}
The last integral can be performed numerically using the 3+1 data, with Cartesian-like coordinates $(x,y)$ parameterized as $\left(x=r\cos\phi,\,y=r\sin\phi\right)$. The introduced auxiliary coordinate $\phi$ might be different from $\varphi$.
This procedure leads us to $g_{\varphi\varphi}(r)= {\mathcal{P}^2}/\left(4\pi^2\right)$.
In addition to $g_{\varphi\varphi}$, the computation of $g_{t\varphi}=g_{\varphi\varphi}\beta^\varphi$ also requires knowledge of $\beta^\varphi$. Taking the square of $\bm{\beta}=\beta^\varphi\,\partial_\varphi$ yields:
\begin{equation*}
\bm{\beta}\cdot{\bm \beta}=\left[\beta^\varphi\right]^2\,g_{\varphi\varphi} \implies \beta^\varphi = \pm \sqrt{\frac{\bm{\beta}\cdot{\bm \beta}}{g_{\varphi\varphi}}}\,.
\end{equation*}
This leads to the relation presented in the main text, $g_{t\varphi}= \pm \sqrt{g_{\varphi\varphi}}\,\sqrt{{\bm \beta}\cdot{\bm \beta}}$. The sign of $g_{t\varphi}$ is determined by the circulation sense of $\bm\beta$ around the coordinate center, being positive (negative) if $\beta^y>0$ ($\beta^y<0$) when $y=0$.

It is helpful to assess how much the spacetime symmetry assumptions are violated. Indeed, one can notice that
\begin{align}
& {\bm \beta} =\beta^x \,\partial_x + \beta^y \,\partial_y =  \underbrace{\beta^r}_{=\,0}\partial_r + \beta^\varphi \,\partial_\varphi\\
& \bm{dr} = \frac{x\,\bm {dx} + y\,\bm{dy}}{r}\\
& \beta^r = \bm{\beta}\bm{dr} = \frac{\beta^x\,x + y\,\beta^y}{r}=0\,.
\end{align}

\begin{widetext}

\begin{figure}
\begin{center}
\includegraphics[width=1\textwidth]{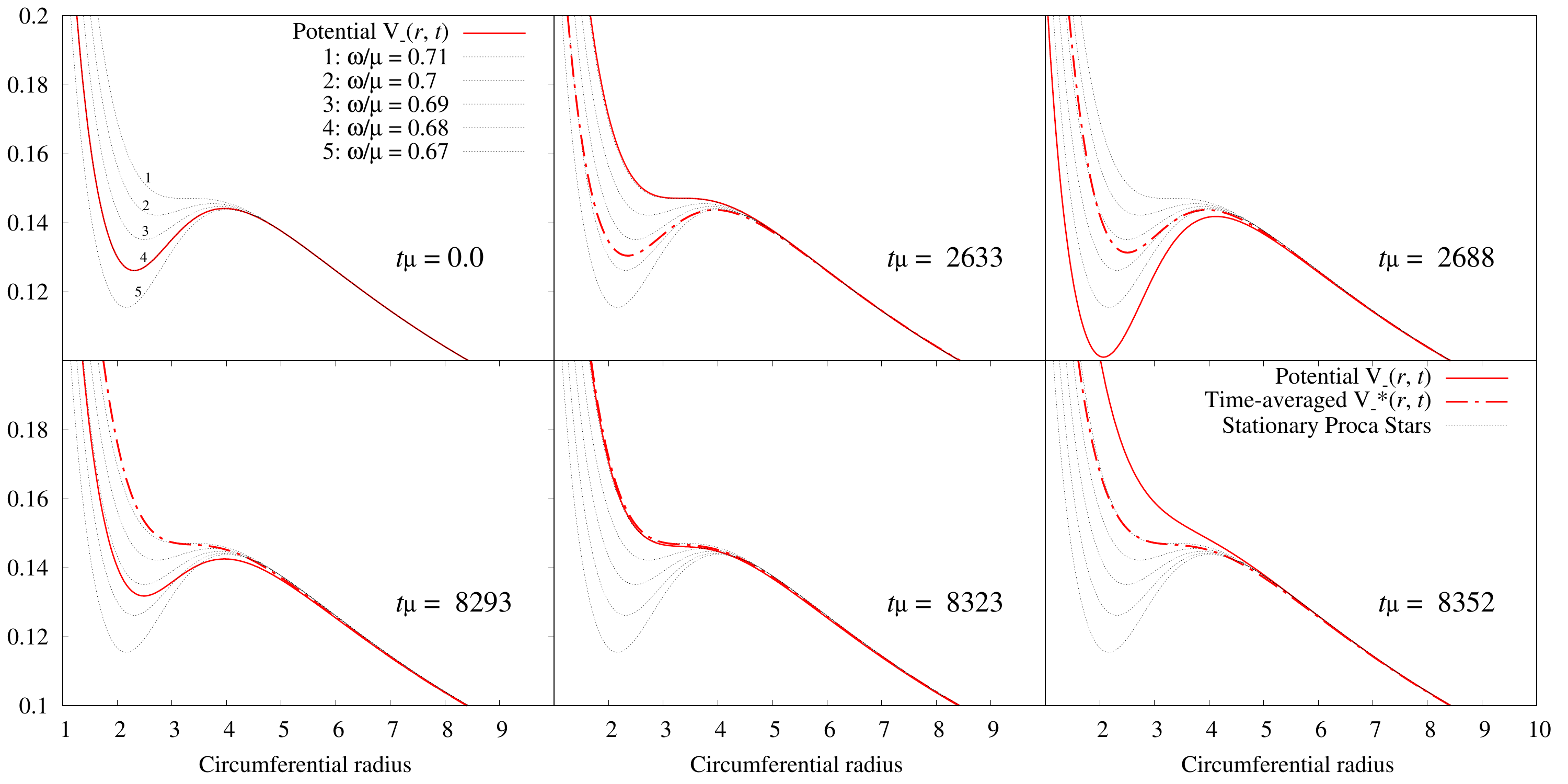}
\caption{\small Show in red are the potentials $V_-(r,t)$ and its time-average $V_-^*(r,t)$ for the dynamical evolution of the Proca Star with $\omega=0.68\mu$, shown at different times $t$. The dashed-grey lines represent the potential of the stationary Proca Star solutions with different $\omega$. The dimensionless circumferential radius $\sqrt{g_{\varphi\varphi}(r)}/M_t$ is used as a geometrical measure of the radial distance to the star center.}
\label{H-plot}
\end{center}
\end{figure}
\end{widetext}

For real numerical relativity data, the quantity $\beta^r$, is not identically zero, and it can be used as a simple measure of the violation of axi-symmetry. For instance, in the Proca Star case departures from symmetry lead to maximum values of $\beta^r\sim 10^{-3}-10^{-2}$. However, these departures from axial-symmetry can be diluted by working with averaged squared functions over surfaces with $r=const.$:
\begin{equation*}
\left<\bm{\beta}\cdot\bm{\beta}\right>\equiv\frac{1}{2\pi}\int_0^{2\pi}\left(\bm{\beta}\cdot\bm{\beta}\right)\,d\phi,\qquad \left<{N^2}\right>\equiv\frac{1}{2\pi}\int_0^{2\pi}N^2\,d\phi\,.
\end{equation*}
Using the latter we define an averaged version of the metric functions:
\begin{align*}
\left<g_{t\varphi}\right> &\equiv \pm \sqrt{g_{\varphi\varphi}}\,\sqrt{\left<{\bm \beta}\cdot{\bm \beta}\right>}\\
\left<g_{tt}\right> &\equiv  -\left<{N^2}\right> + \left<{\bm \beta}\cdot{\bm \beta}\right>\\
\left<g_{\varphi\varphi}\right> &\equiv g_{\varphi\varphi}\,,
\end{align*}
with the AEP taking the explicit form:
\begin{equation}
V_\pm(r,t)  = M_t\,\frac{\left<g_{t\varphi}\right>\mp\sqrt{\left<g_{t\varphi}\right>^2-\left<g_{tt}\right>\left<g_{\varphi\varphi}\right>}}{\left<g_{\varphi\varphi}\right>}\,.
\end{equation}
This dimensionless potential can be defined for each time slice $t$, assuming an adiabatic evolution of the spacetime. Profiles of $V_\pm$ taken at different times can be compared directly, with possible mass loss during the evolution already accounted for.

As a concrete spacetime example, we focus on Proca Star evolutions described in the main text. Close to the end of their evolution, Proca Stars can be considered in a perturbed excited state, with metric functions ($i.e.$ the lapse) oscillating periodically around an average behaviour. Some long-term patterns, such as the slow merger of LRs during the evolution, become much clearer in terms of the time-averaged potential $V_\pm^*$, defined in the main text, rather than the original potential $V_\pm(r,t)$. The fact that a typical Proca Star oscillation period is around $\sim 100$ times larger than the light-crossing time between the LRs further validates the adiabatic approximation. For the Proca Stars considered here, the LRs appear only when selecting the negative sign of the potential, $i.e.$ for $V^*_-$.

The evolution of the potential $V_-(r,t)$ and its time-average $V_-^*(r,t)$ are represented in Fig.~\ref{H-plot} for the Proca Star with $\omega=0.68\mu$. We can notice that even for large $t$, the profile of the dynamical potentials closely resemble those of stationary Proca Star solutions with different $\omega$. This might suggest that dynamical Proca Stars closely match the structure of different stationary solutions during their time evolution, although this interpretation is likely a naive approximation. Nevertheless, the consistency of the AEP profile with that of several stationary Proca Stars, independently of the numerical relativity time evolution, is a further indication on the validity of the AEP framework.

The bottom row of Fig.~\ref{H-plot} displays three snapshots of the potentials during a single oscillation of the star. The potential $V_-(r,t)$ possesses two LRs at $t\mu=8293$, which are then destroyed as we cross $t\mu=8323$ and $t\mu=8352$. In contrast, the time-averaged potential $V^*(r,t)$ remains fairly unchanged in the course of a single oscillation of the star. Ultimately, the purpose of $V_-^*$ is to represent the unexcited reference state of the star at that time.

\bigskip

\begin{figure}[t!]
\centering
\includegraphics[width=0.245\linewidth]{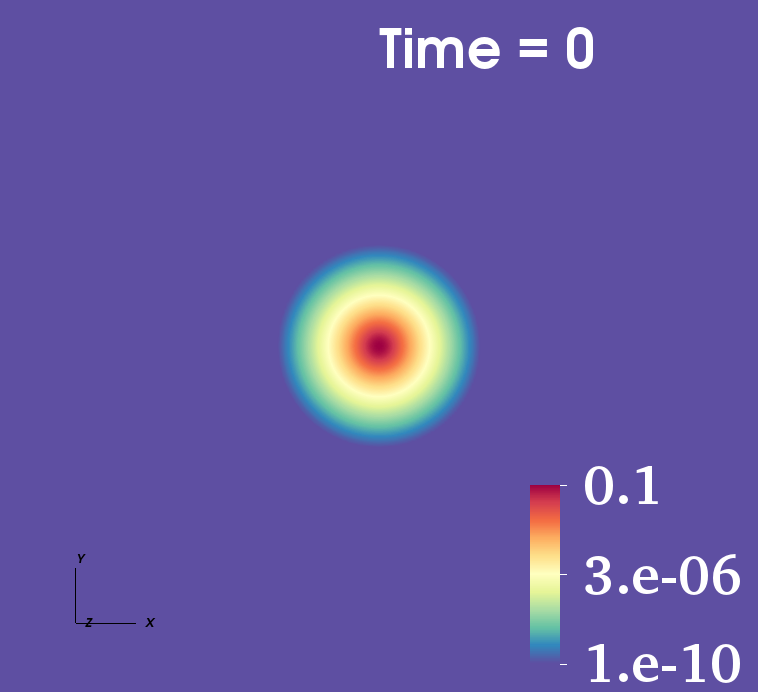}\hspace{-0.01\linewidth}
\includegraphics[width=0.245\linewidth]{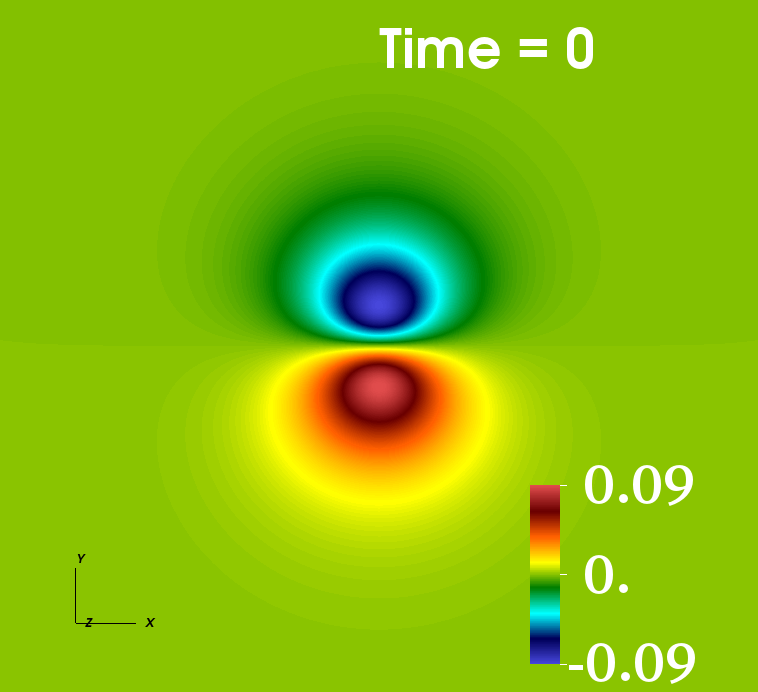}\hspace{-0.01\linewidth}
\includegraphics[width=0.245\linewidth]{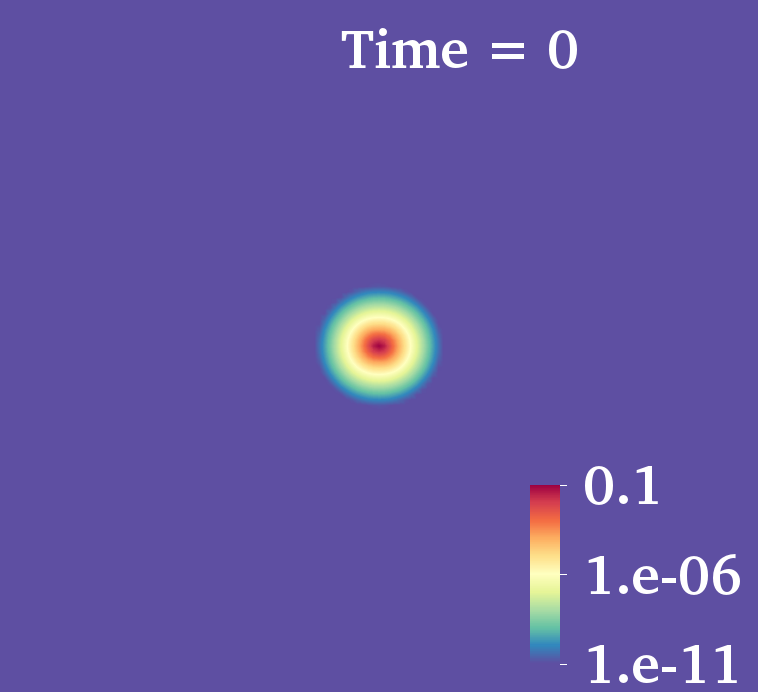}\hspace{-0.01\linewidth}
\includegraphics[width=0.245\linewidth]{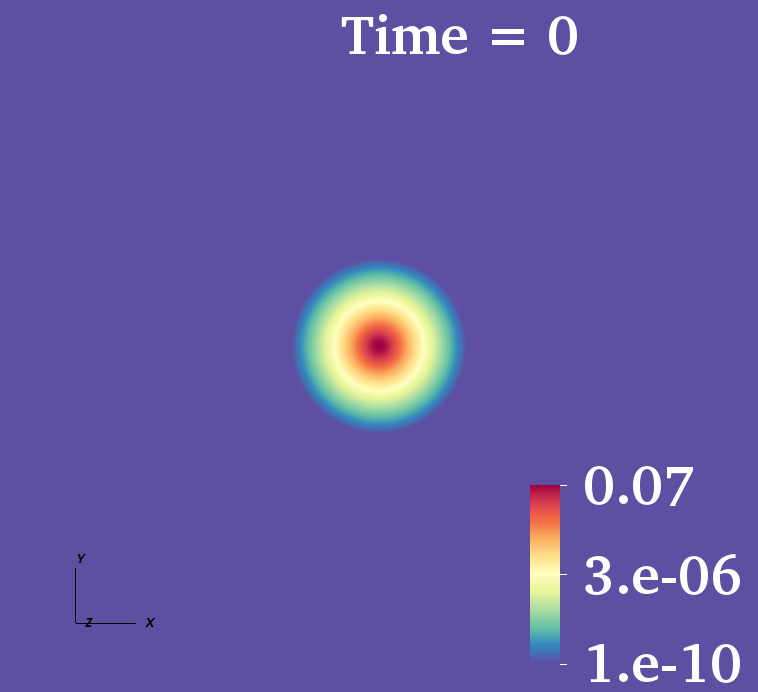}\hspace{-0.01\linewidth}\\
\includegraphics[width=0.245\linewidth]{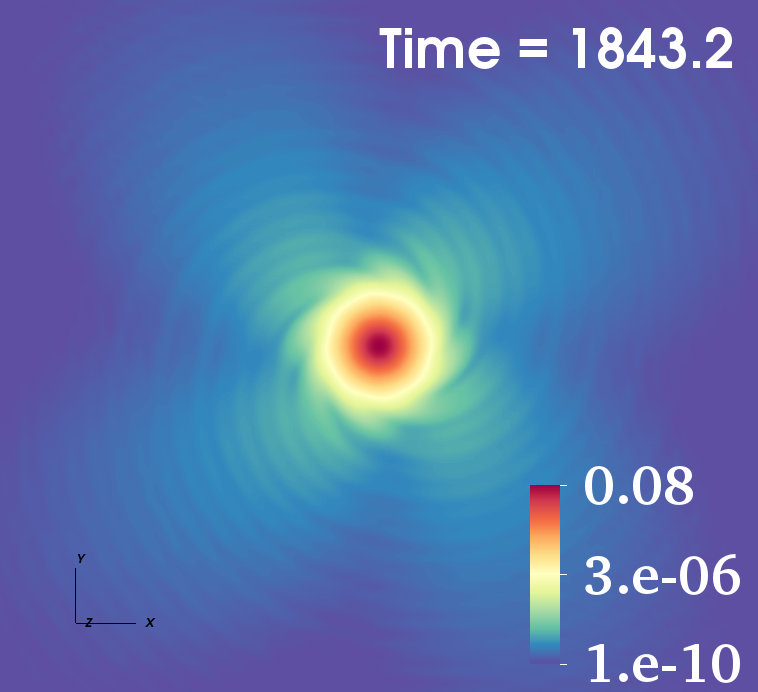}\hspace{-0.01\linewidth}
\includegraphics[width=0.245\linewidth]{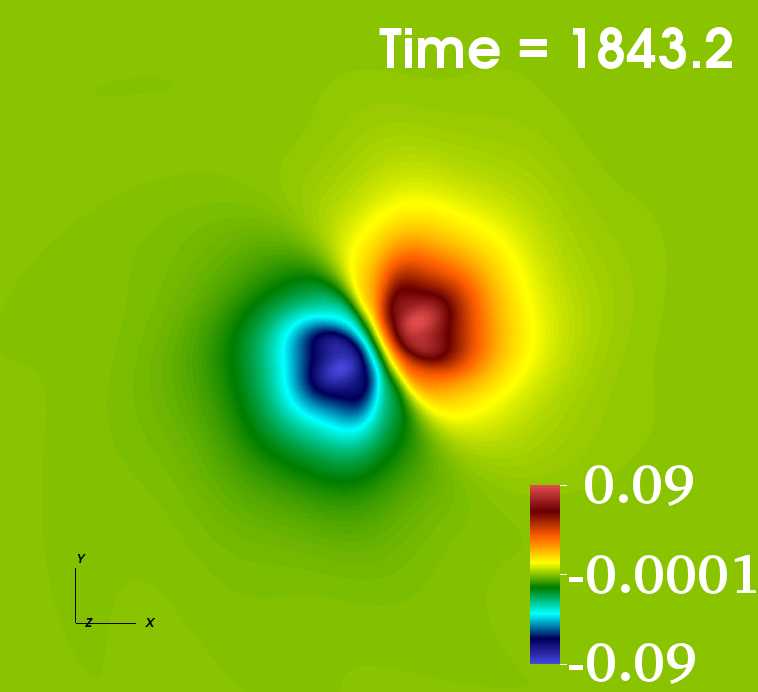}\hspace{-0.01\linewidth}
\includegraphics[width=0.245\linewidth]{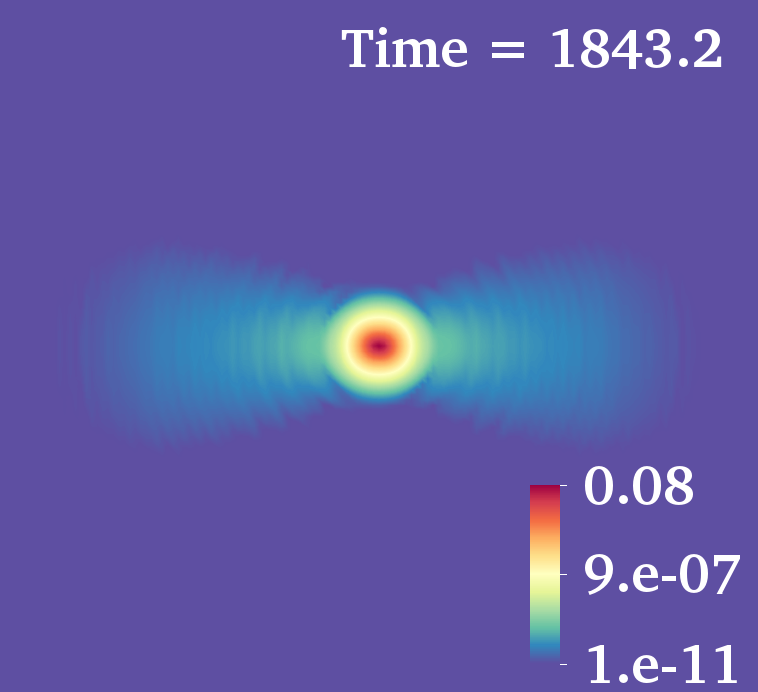}\hspace{-0.01\linewidth}
\includegraphics[width=0.245\linewidth]{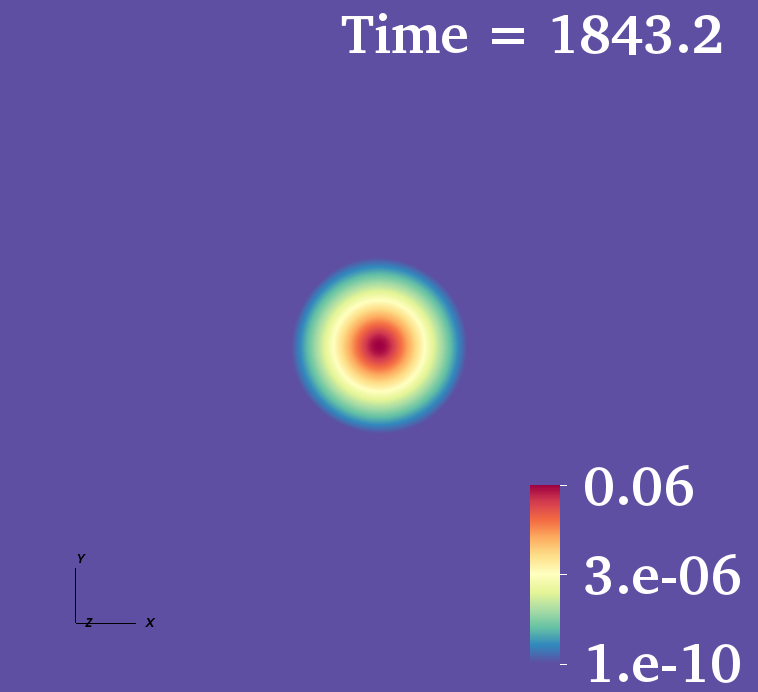}\hspace{-0.01\linewidth}\\
\includegraphics[width=0.245\linewidth]{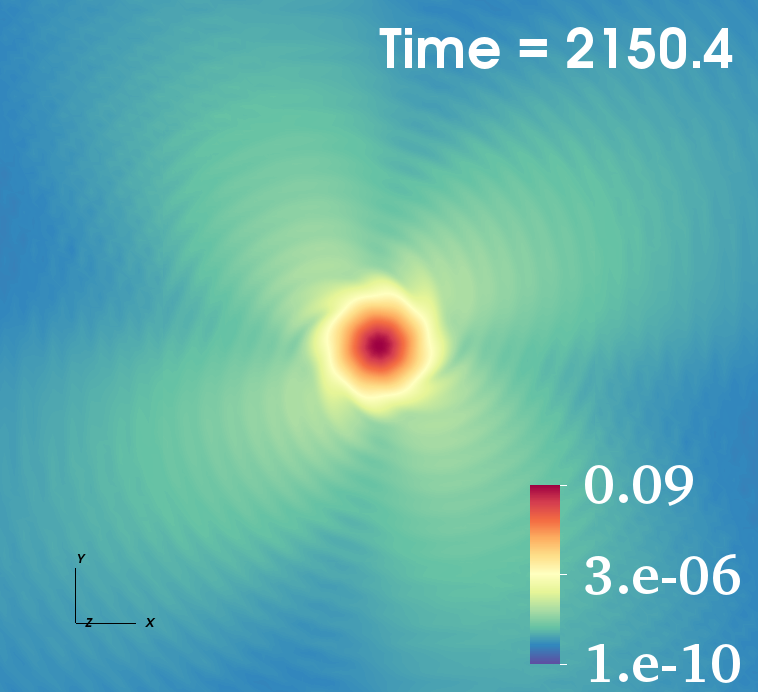}\hspace{-0.01\linewidth}
\includegraphics[width=0.245\linewidth]{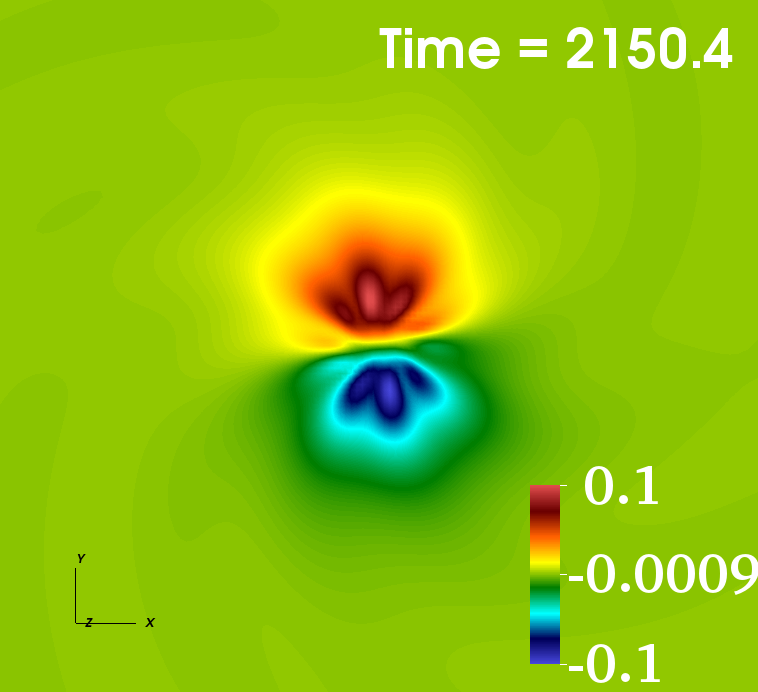}\hspace{-0.01\linewidth}
\includegraphics[width=0.245\linewidth]{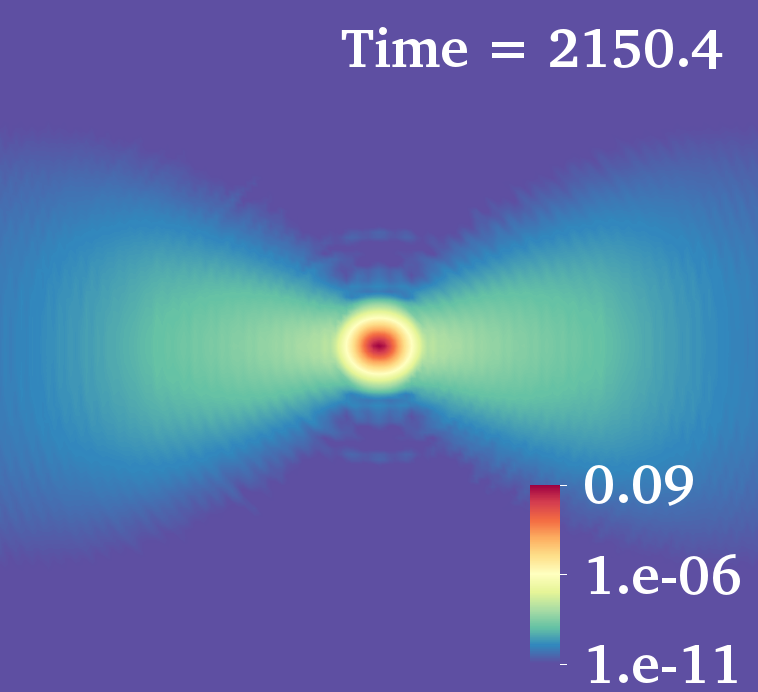}\hspace{-0.01\linewidth}
\includegraphics[width=0.245\linewidth]{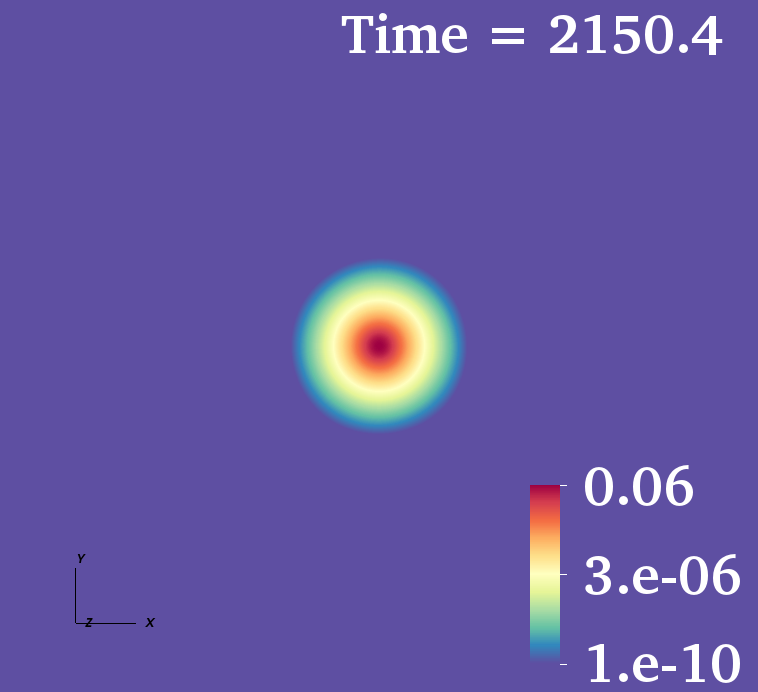}\hspace{-0.01\linewidth}\\
\includegraphics[width=0.245\linewidth]{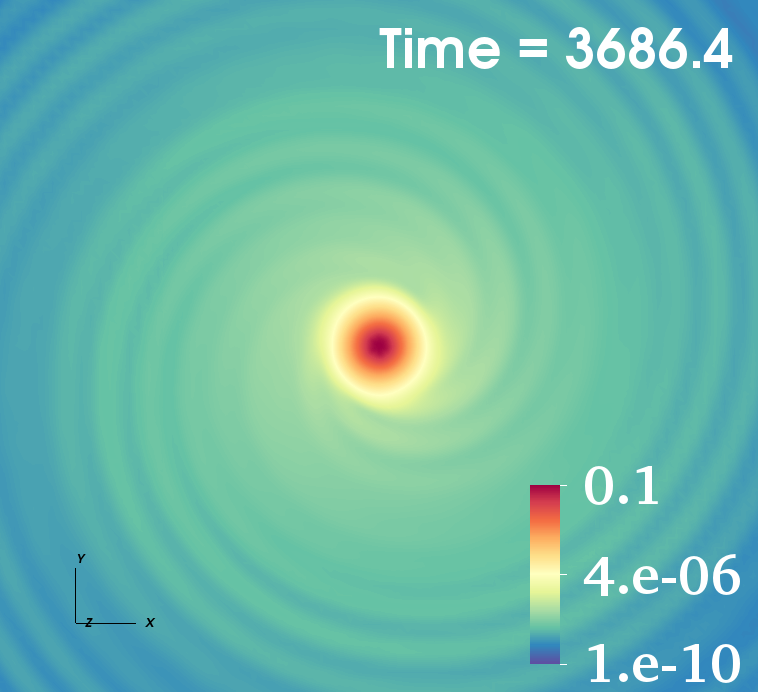}\hspace{-0.01\linewidth}
\includegraphics[width=0.245\linewidth]{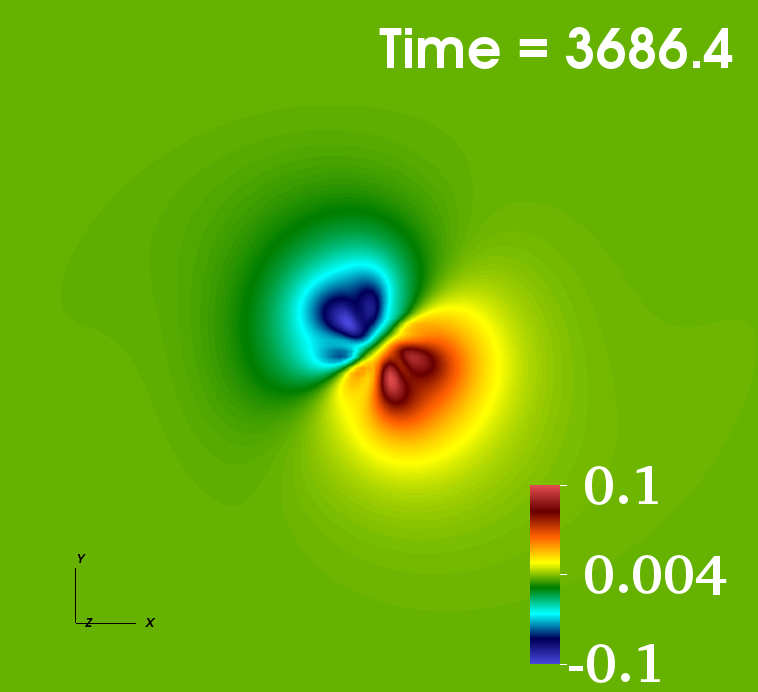}\hspace{-0.01\linewidth}
\includegraphics[width=0.245\linewidth]{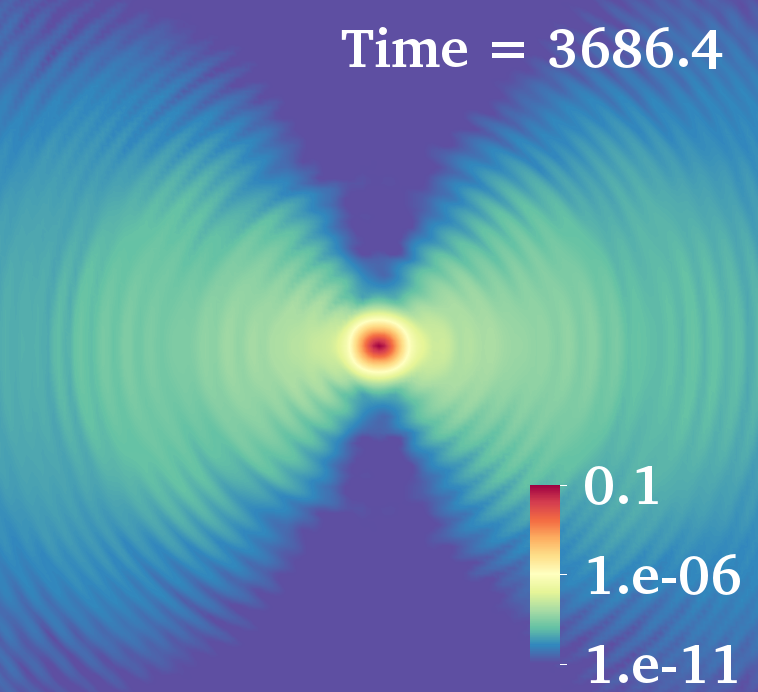}\hspace{-0.01\linewidth}
\includegraphics[width=0.245\linewidth]{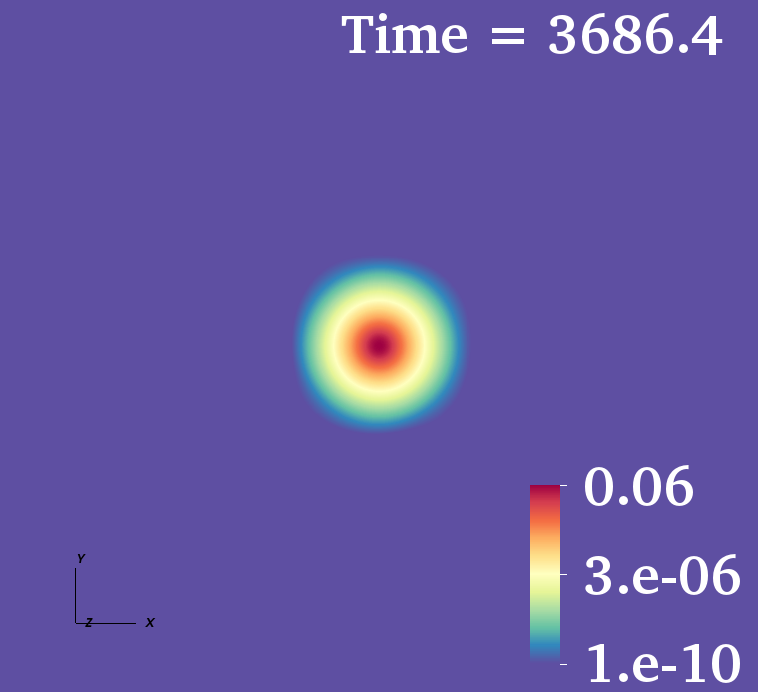}\hspace{-0.01\linewidth}\\
\includegraphics[width=0.245\linewidth]{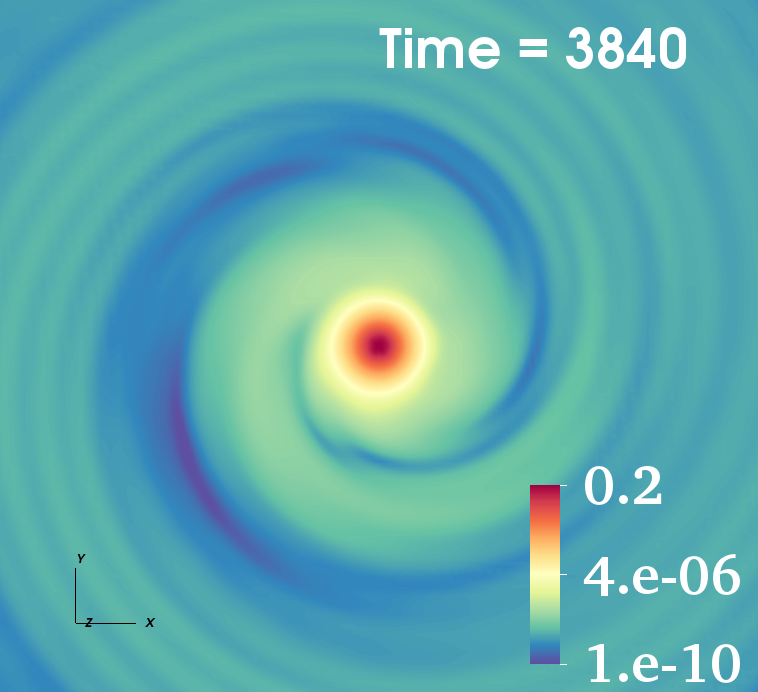}\hspace{-0.01\linewidth}
\includegraphics[width=0.245\linewidth]{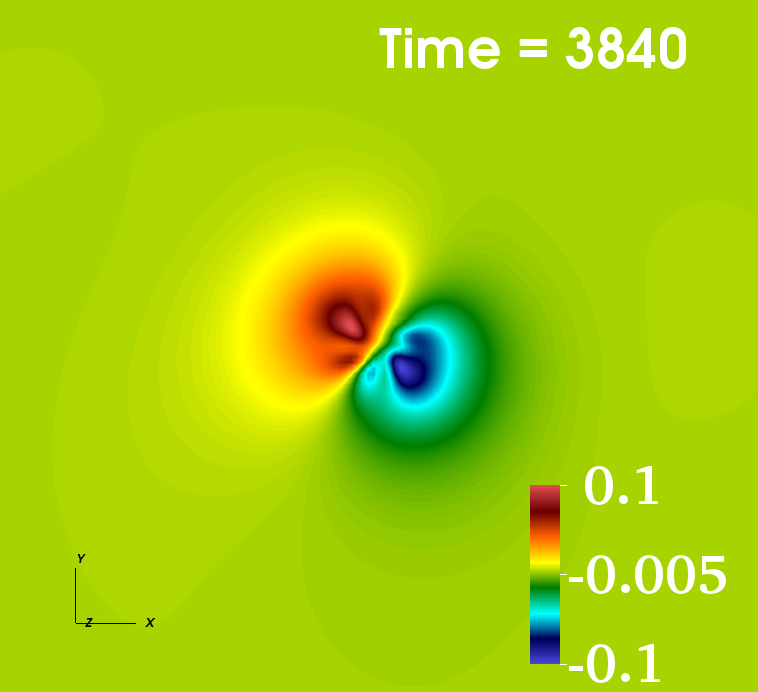}\hspace{-0.01\linewidth}
\includegraphics[width=0.245\linewidth]{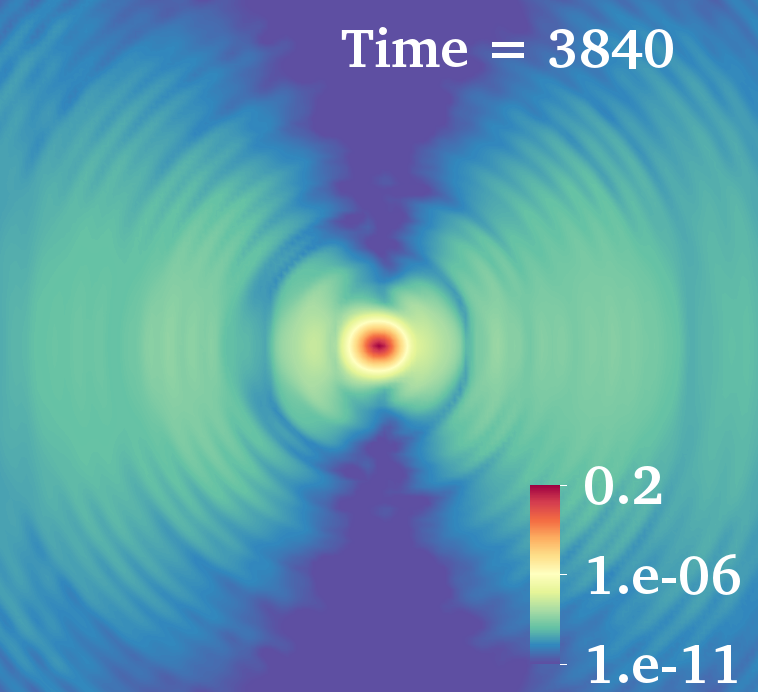}\hspace{-0.01\linewidth}
\includegraphics[width=0.245\linewidth]{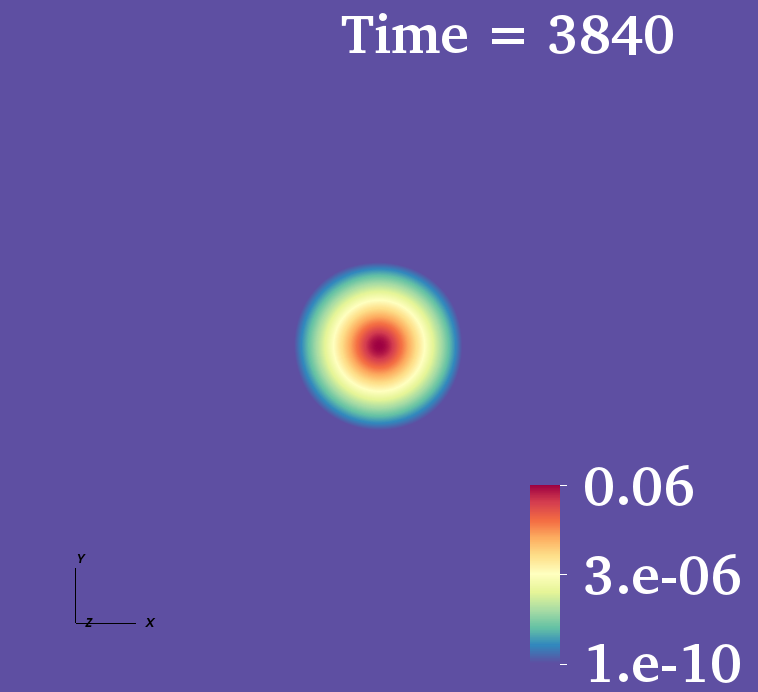}\hspace{-0.01\linewidth}\\
\includegraphics[width=0.245\linewidth]{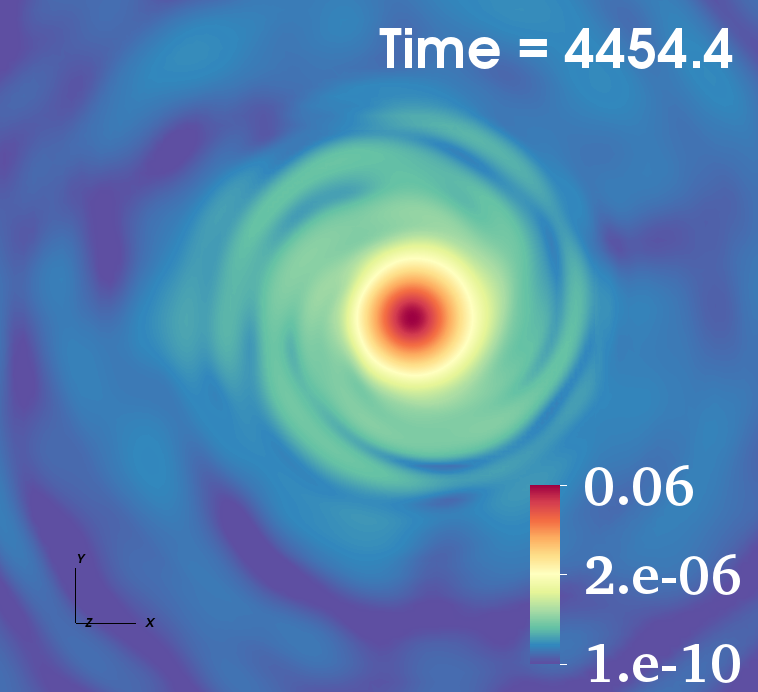}\hspace{-0.01\linewidth}
\includegraphics[width=0.245\linewidth]{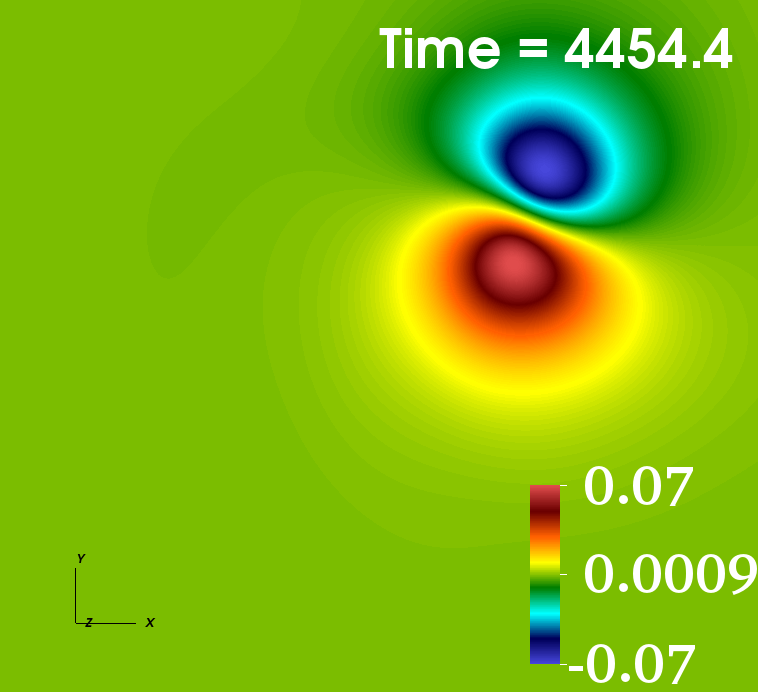}\hspace{-0.01\linewidth}
\includegraphics[width=0.245\linewidth]{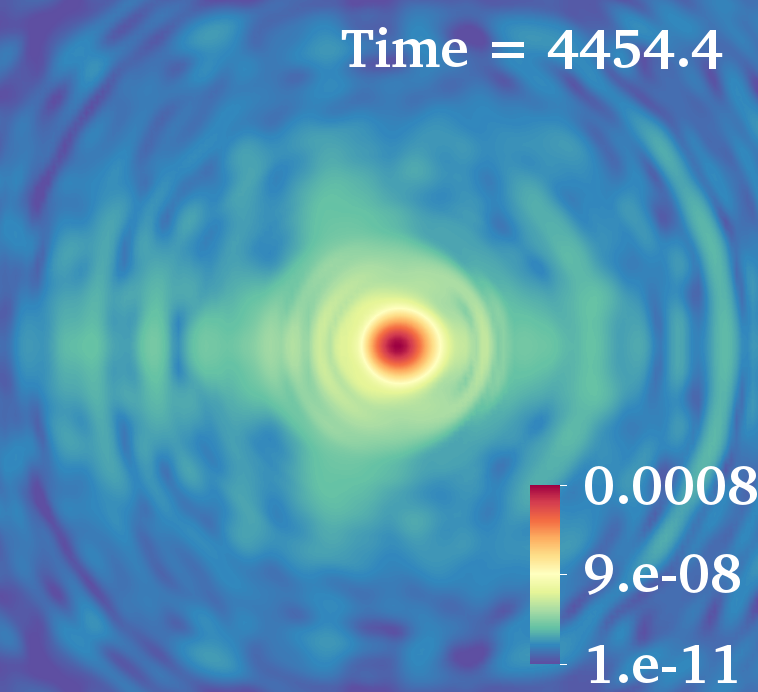}\hspace{-0.01\linewidth}
\includegraphics[width=0.245\linewidth]{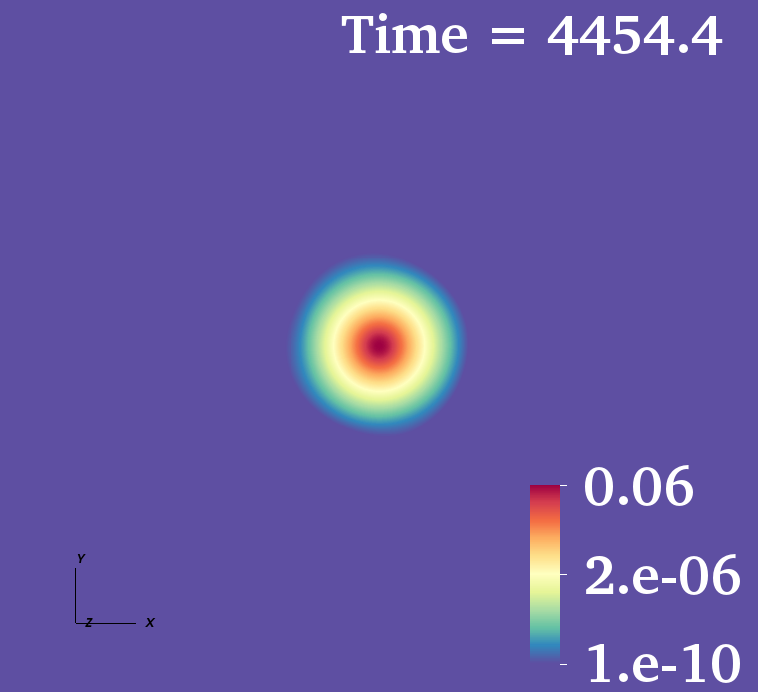}\hspace{-0.01\linewidth}\\
\includegraphics[width=0.245\linewidth]{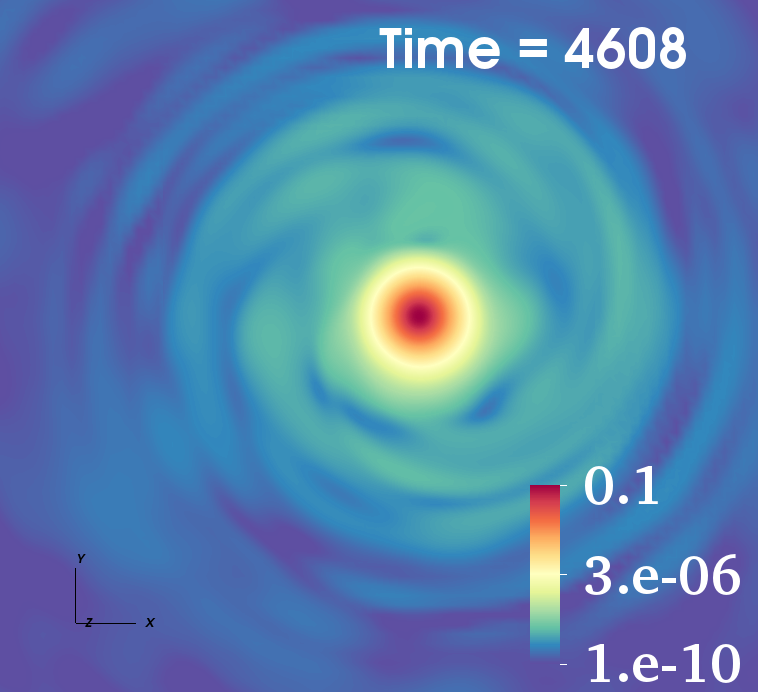}\hspace{-0.01\linewidth}
\includegraphics[width=0.245\linewidth]{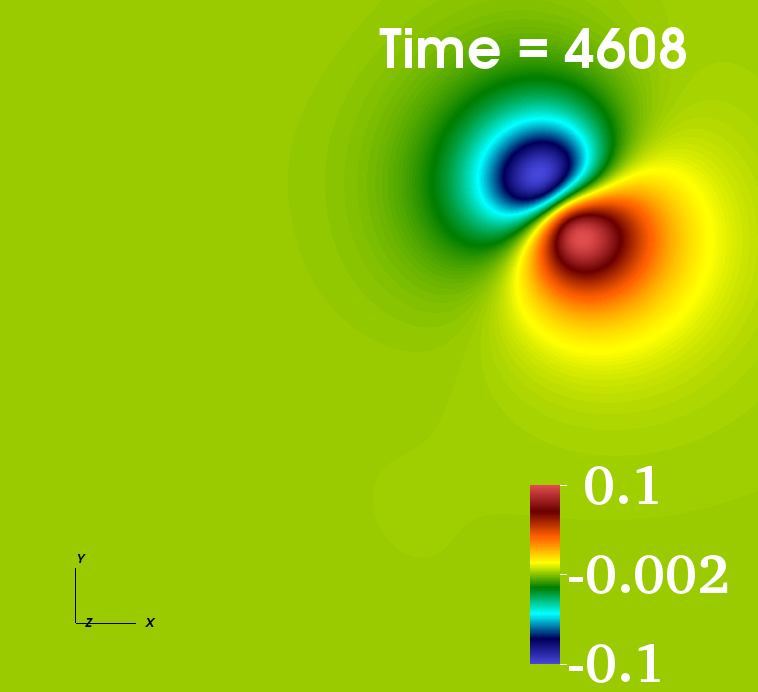}\hspace{-0.01\linewidth}
\includegraphics[width=0.245\linewidth]{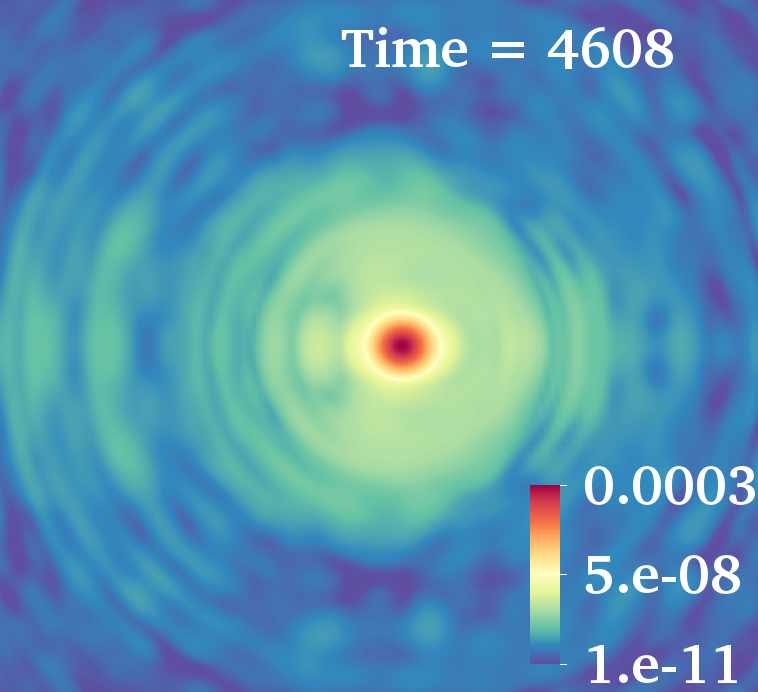}\hspace{-0.01\linewidth}
\includegraphics[width=0.245\linewidth]{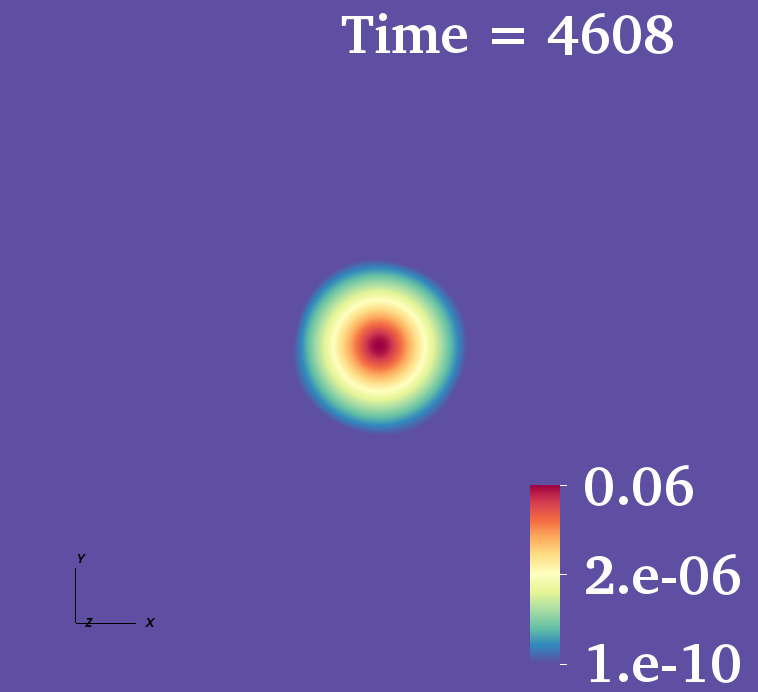}\hspace{-0.01\linewidth}\\
\includegraphics[width=0.245\linewidth]{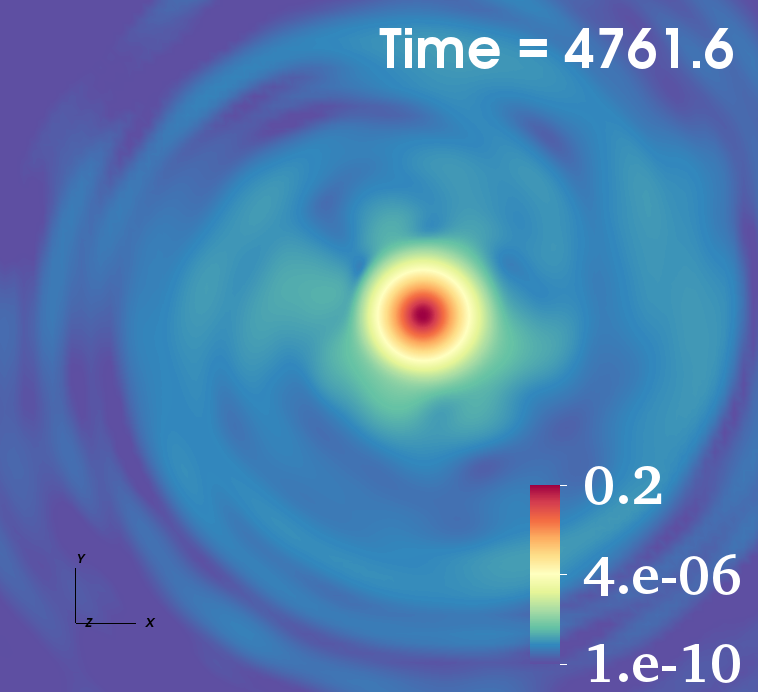}\hspace{-0.01\linewidth}
\includegraphics[width=0.245\linewidth]{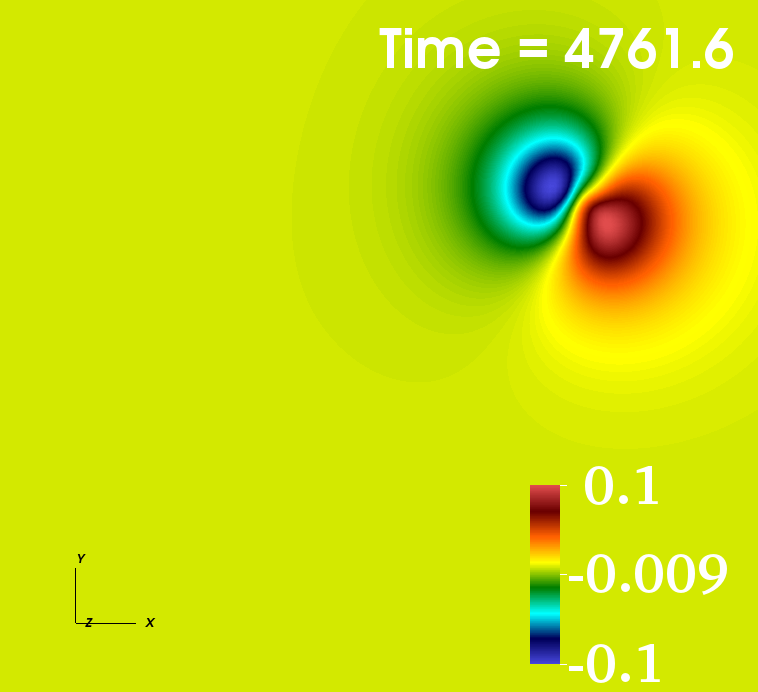}\hspace{-0.01\linewidth}
\includegraphics[width=0.245\linewidth]{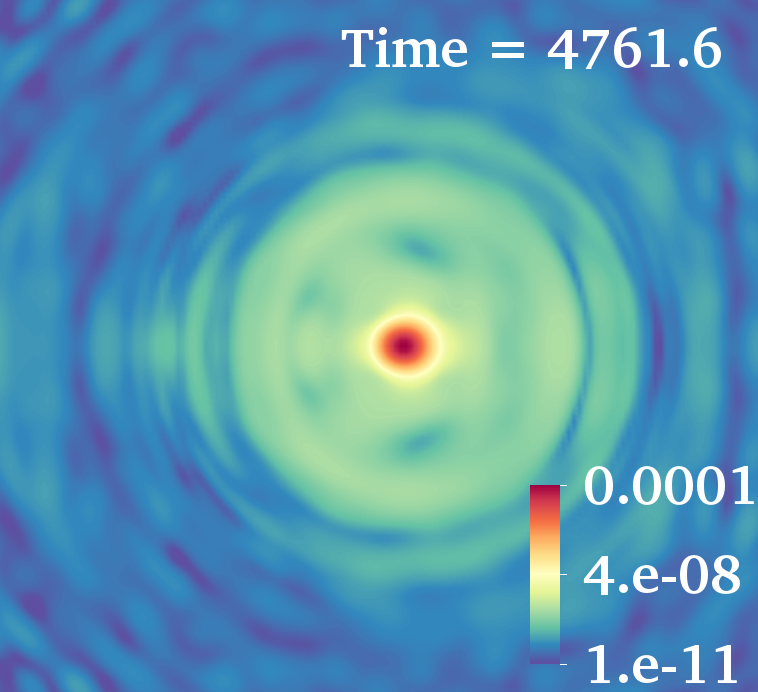}\hspace{-0.01\linewidth}
\includegraphics[width=0.245\linewidth]{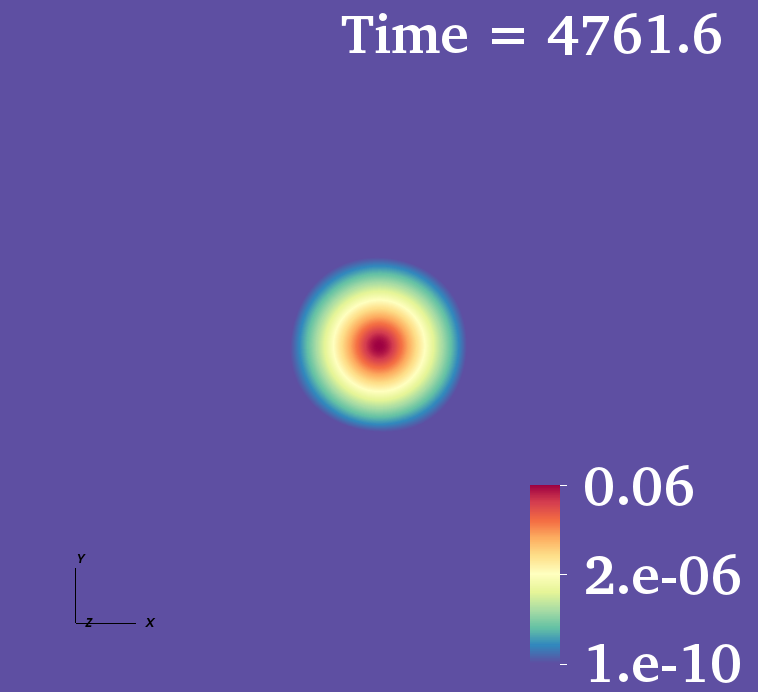}\hspace{-0.01\linewidth}\\
\caption{Time evolution of an ultracompact spinning Proca Star with  $\omega/\mu=0.70$: energy density in logscale (first column) and the real part of the scalar potential (second column) both on the equatorial plane, and the energy density in logscale in the $xz$ plane (third column). For comparison, the energy density in logscale of a stable model with $\omega/\mu=0.72$ (fourth column).}
\label{figLogproca}
\end{figure}

{\bf {\em Appendix B. Loss of energy and angular momentum in the migration scenario.}} 
The loss of Proca field during the instability is shown in Fig.~\ref{figLogproca} where we plot the energy density in logscale on the equatorial plane and the $xz$ ($y=0$) plane of the Proca star with $\omega/\mu=0.70$. The energy (and angular momentum) densities are computed from Komar integrals as in~\cite{Cunha:2017wao,Sanchis-Gual:2018oui}. The Proca field is emitted mainly along the equatorial plane, with some of it falling back onto the star, yielding a breathing behaviour. Such loss is not present in the scalar solitonic bosonic stars studied, and may play an important role in the different outcomes. The real part of the scalar potential $\mathcal{X}_{\phi}$ of the Proca field (see~\cite{Sanchis-Gual:2018oui} for the definition of this potential) is shown in the second column of Fig.~\ref{figLogproca}; it illustrates how the instability affects the star at the level of the Proca field (see the panels in the third, fourth, and fifth rows). The ``fleur-de-lys" pattern in the Proca field is what configures the polynomial features in the energy density. The bottom middle panel proves that the end-point of the instability is still an $m=1$ spinning Proca star.

\begin{figure}[h!]
\centering
\includegraphics[width=1.0\linewidth]{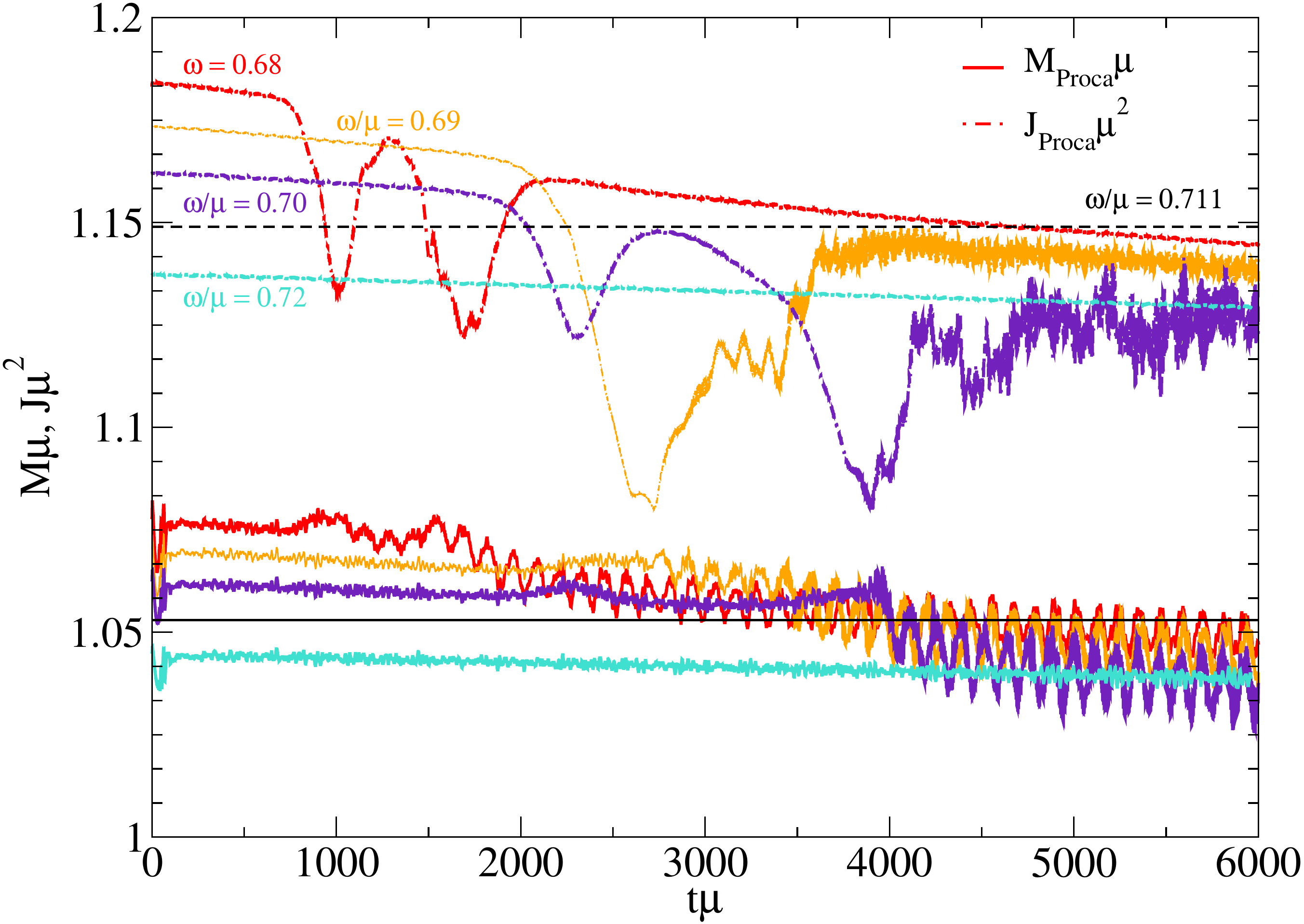}
\caption{Time evolution of the mass and angular momentum of four Proca star models. For comparison, the critical Proca star, for which LRs first appear has $\omega/\mu\simeq 0.711$, $ M\mu\simeq 1.053$ (horizontal black solid line) and $J\mu^2\simeq 1.149$ (horizontal black dashed line). It is clear that the evolution of the ultracompact Proca stars tends to migrate them to non-ultracompact stars.}
\label{figEnergy}
\end{figure}

Fig.~\ref{figEnergy} exhibits the time evolution of the mass and angular momentum of four spinning Proca stars with $\omega/\mu=0.68$, 0.69, 0.70 (with LR), and 0.72 (without LR). The models with LR suffer an important loss of angular momentum when the instability kicks in, but part of the angular momentum is accreted back by the Proca star. After an initial more violent process of the instability, the stars continue to lose angular momentum and mass at a slower rate,  through gravitational cooling,  tending towards values corresponding to configurations without LR. This is the part wherein we expect the AEP to be a good physical description of the LRs' evolution. On the other hand, as expected, the model with $\omega/\mu=0.72$ does not show any instability. In the latter a drift is seen in the conservation of the total energy and angular momentum due to numerical error, but this drift goes away with resolution~\cite{sanchis2019nonlinear}. The deviation at the end of the simulation with respect to the initial values is less than 1\%. In the bottom panels of Fig.~\ref{figLogproca} we can see that the gravitational cooling is still at work and there is a physical ejection of material, which is not present in the stable case.

\begin{figure}[h!]
\centering
\includegraphics[width=1.0\linewidth]{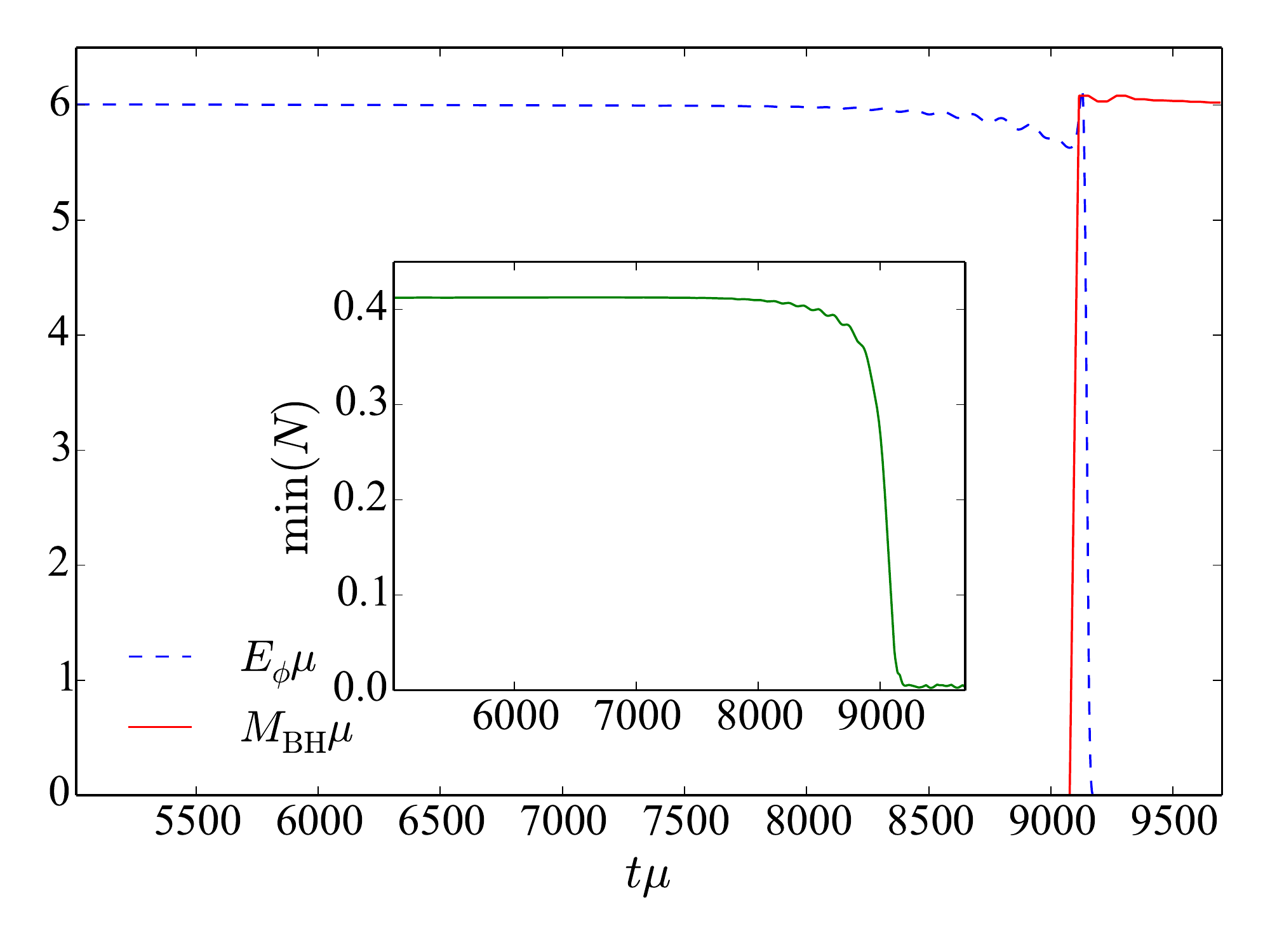}
\caption{Time evolution of the scalar field energy $E_\varphi$ and BH mass $M_{\rm BH}$ for the ultracompact solitonic spinning scalar boson star with $\omega/\mu=0.16$, which collapses into a BH.}
\label{figBH}
\end{figure}

Finally, in Fig.~\ref{figBH} we plot the BH formation after the collapse of the spinning solitonic scalar boson star with $\omega/\mu=0.16$. The inset shows the minimum value of the lapse function $N$ that approximately vanishes when the apparent horizon forms.

\bibliography{num-rel2}

\end{document}